\documentclass[11pt]{article}

\usepackage[margin=1in]{geometry}
\usepackage{times}
\usepackage[utf8]{inputenc}

\usepackage{amsmath}
\usepackage{amssymb}
\usepackage{bm}
\usepackage{mathtools}
\usepackage{graphicx}
\usepackage{booktabs}
\usepackage{enumitem}
\usepackage{float}
\usepackage{multirow}
\usepackage{caption}
\usepackage{subcaption}
\usepackage{xcolor}
\usepackage[section]{placeins}
\usepackage{natbib}
\usepackage{url}
\usepackage{hyperref}

\numberwithin{equation}{section}

\newcommand{\bp}[1]{\mathbf{#1\%}}
\newcommand{\mc}[1]{\mathcal{#1}}

\title{Learning Nonlinear Transient Dynamics in Fluid Flow Networks with the Neural Basis Method}

\author{%
Aijaz Nazir\thanks{Dept.\ of Mathematics, University of Houston, Houston, TX 77204, ORCID: 0000-0002-3494-4101, anazir@cougarnet.uh.edu}
\and
Yuhe Wang\thanks{Asia Pacific Technology, Inc., Houston, TX 77042, USA; Institute for Scientific Computation, Texas A\&M University, College Station, TX 77843, USA. \texttt{yuhe.wang@me.com}}
\and
S.K. Rasel \thanks{Department of Mathematics, University of Houston, Houston, TX 77204, USA. \texttt{skrasel@cougarnet.uh.edu}}
\and
Min Wang\thanks{Department of Mathematics, University of Houston, Houston, TX 77204, USA. \texttt{mwang55@central.uh.edu}}
}

\date{\today}

\begin{document}

\maketitle

\begin{abstract}
Fluid flow networks are ubiquitous across engineered and natural systems, including pipeline infrastructure and biological vasculature.
In many practical regimes, they are governed by nonlinear edgewise hyperbolic laws coupled through continuity and conservation at junctions.
Learning their dynamics is important in many-query scenarios requiring rapid evaluation across unseen parametric settings. 
Conventional methods re-solve the system for each new query, resulting in prohibitive cost. 
Existing physics-informed neural approaches typically impose governing equations through penalty losses over generic neural approximators, rather than embedding the physical laws in their neural formulation.
Such losses act only as optimization surrogates during training, with no 
guarantee that their reduction reduces solution error. 
Moreover, the resulting optimization problem is prone to ill-conditioning, particularly for hyperbolic systems.
Building on the ``representation-then-projection'' principle of the neural basis method, we introduce a new approach for modeling fluid flow networks. 
Fixed neural bases define the representation space on each edge, while projection of the discretized equations determines the expansion coefficients, embedding edgewise laws into a conservative formulation.
Edge subdomains are coupled through non-overlapping domain-decomposition.
A neural operator then learns, directly in this reduced space, the parametric map to the corresponding expansion coefficients.
On two representative network topologies, looped and branching, we reproduce references
with relative $L^2$ errors below $0.02\%$, whereas the physics-informed neural network baseline fails. 
Across unseen boundary conditions in both interpolation and extrapolation, we maintain mean pressure errors below $1\%$ and mean flux errors below $3.3\%$, while the physics-informed DeepONet baseline reaches $8$--$18\%$ pressure errors and flux errors of $85$--$95\%$, failing to capture the underlying dynamics. 
The results demonstrate the accuracy and efficiency of the proposed framework for learning nonlinear transient dynamics in fluid flow networks, highlighting its potential for real-time monitoring, design exploration, and uncertainty quantification.

\end{abstract}

\noindent\textbf{Keywords:}
fluid flow network, nonlinear transient dynamics, operator learning, physics-informed machine learning, neural basis method

\section{Introduction}
\label{sec:intro}

Fluid flow networks, such as natural gas transmission systems, are
governed by nonlinear hyperbolic conservation laws on each pipe,
coupled at junctions by pressure continuity and mass conservation
\citep{banda2006gas}. Planning, control, and uncertainty quantification
require the transient network state under many boundary conditions,
and re-solving with conventional methods for each query is the bottleneck.
Current learned surrogates face critical limitations:
physics-informed networks and PI-DeepONet enforce the equations as
heuristically weighted penalty losses whose magnitude is not tied to
solution error, and which are poorly conditioned for transient
hyperbolic dynamics \citep{wang2021learning, strelow2023physics};
data-driven neural operators do not embed junction coupling or
discrete conservation \citep{lu2019deeponet, li2020fourier}. The Neural
Basis Method (NBM) \citep{wang2026solving} avoids penalty training by
projecting discretized equations onto a fixed neural basis, but has so
far been developed with global bases on single domains. Networks, however, demand a representation that is local to each pipe yet conserves mass across junctions, and a discretization that stays stable for nonlinear hyperbolic transients.

We introduce NBM to fluid flow networks. Each edge carries a local neural
basis, junction pressures enter as dedicated coefficients, and the
staggered finite-volume equations of the whole network are projected
in a single monolithic least-squares solve. A network trained on the
projected discrete residual then maps time and inlet pressure to a
POD-compressed latent state, returning the transient at any queried
time without time stepping.

Our contributions are summarized as:
\begin{itemize}[leftmargin=*, itemsep=1pt]
\item \textbf{Localized NBM on metric graphs.} We couple edgewise
bases through junction conditions in one monolithic projection,
extending NBM from single-domain global bases to network domains.
\item \textbf{NBM for nonlinear hyperbolic systems.} Unlike
collocation on the strong form, we project a
staggered finite-volume discretization of the isothermal Euler
equations onto the neural basis, extending NBM beyond the
elliptic--parabolic setting in which it was introduced; it recovers the discrete solution on both looped and branching topologies to relative $L^2$ errors of
$10^{-5}$ in pressure and $10^{-4}$ in flux.
\item \textbf{Operator learning on networks.} NBM-OL predicts the
transient state at held-out inlet pressures in a single forward pass.
Averaged over held-out points, relative $L^2$ errors in
pressure/flux are $0.065\%/1.05\%$ (interpolation) and
$0.23\%/1.62\%$ (extrapolation) on the 7-edge network, and
$0.12\%/3.25\%$ and $0.97\%/2.84\%$ on the tree network, compared with
$7.9$--$17.9\%$ pressure and $85$--$95\%$ flux errors for PI-DeepONet.
\end{itemize}

\section{Related Work}
\label{sec:related}

\textbf{Neural operators.} Neural operators learn maps between function
spaces that approximate the solution operator of a PDE
\citep{kovachki2023neural}. DeepONet factorizes the operator into
branch and trunk networks \citep{lu2019deeponet}, and the Fourier
Neural Operator parameterizes the kernel in the Fourier domain
\citep{li2020fourier}. PI-DeepONet adds the PDE residual to the
training objective, reducing reliance on paired data
\citep{wang2021learning}; we use a graph-adapted variant as a baseline.
Large pretrained models such as MPP \citep{mccabe2024multiple}, DPOT
\citep{hao2024dpot}, and Poseidon \citep{herde2024poseidon} typically
learn a one-step update applied autoregressively, which can accumulate
error over long rollouts \citep{lippe2023pde}; continuous-time
alternatives learn the dynamics directly \citep{hou2026cfo}. Closest to
our setting, \citet{wu2025physics} and \citet{vzdrale2026hypno} develop
neural operators for networked systems and for hyperbolic conservation
laws on graphs. NBM-OL learns a direct map, here
from $(t, P_{\mathrm{in}})$ to the latent state, but represents the
solution in a fixed basis obtained by projecting the discretized
network equations rather than in a learned one.

\textbf{Fixed neural bases and least-squares projection.} A separate
line of work fixes the network weights and solves for expansion
coefficients by least squares, rather than training a network to
satisfy the PDE. Random feature methods \citep{chen2022bridging} and
extreme learning machines use randomly initialized single-hidden-layer
networks as basis functions; local variants partition the domain and
couple subdomains through interface conditions
\citep{dong2021local}. These constructions are effective but their
non-orthogonal bases can become severely ill-conditioned as the basis
is enriched. The Neural Basis Method \citep{wang2026solving} addresses
this with a multilayer basis generator, physics-conforming vector
bases, and residual weighting consistent with the operator, and
extends the construction to parametric operator learning. It has so
far been developed with global bases on single domains, with
localization identified as an open direction. We take that step for
network domains: edgewise bases coupled through junction rows, applied
to a nonlinear hyperbolic system, with the projection performed on a
staggered finite-volume discretization rather than on collocated
strong-form residuals.

\textbf{Reduced-order and latent-space operator learning.} Projection-based
reduced-order models classically compress the state on a POD subspace
and recover the reduced dynamics either by Galerkin projection or, as
in least-squares Petrov--Galerkin schemes, by minimizing the discrete
residual over the reduced space \citep{carlberg2017galerkin}. POD-NN
replaces the reduced solve with a network regressing POD coefficients
on the parameters \citep{hesthaven2018non}, and
\citet{kontolati2024learning} learn autoencoder latents and a neural
operator evolving them. NBM-OL sits between these: the spatial
reduction is a fixed, untrained basis rather than a learned encoder,
the latent dynamics are a direct map from time and parameter rather
than a reduced time-stepping scheme, and training minimizes the
projected discrete residual, keeping the reduced coordinate
interpretable in terms of the finite-volume discretization.

\textbf{Numerical methods for gas pipeline networks.} Transient gas
flow in pipeline networks is classically formulated as the isothermal
Euler equations on a metric graph with Kirchhoff-type coupling at
junctions \citep{osiadacz1984simulation, banda2006gas,
banda2006coupling, herty2008coupling}. Explicit staggered-grid schemes
conserve mass exactly by construction \citep{gyrya2019explicit}, and
operator-splitting schemes give unconditionally stable, second-order
simulation on looped networks \citep{dyachenko2017operator}.
\citet{strelow2023physics} study PINNs for gas transport and find that
adaptive loss balancing offers no advantage over well-chosen constant
weights. Our staggered finite-volume discretization
(Appendix~\ref{sec:governing}) follows the same physical formulation,
but serves as the target of a reduced, learnable projection rather than
as a standalone forward solver.

\section{Problem Setup}
\label{sec:setup}

\paragraph{Network model.}
We consider isothermal compressible gas flow on a pipeline network,
represented as a directed metric graph $\mc{G} = (\mc{V}, \mc{E})$. Each edge
$e \in \mc{E}$ is identified with an interval $[0, L_e]$, oriented from its
tail vertex ($x = 0$) to its head vertex ($x = L_e$); the orientation
fixes the sign convention for flow. On each edge the state consists of
the pressure $P_e(x,t)$ and the mass flux $q_e(x,t) = \rho_e u_e$
(mass flow per unit cross-sectional area), which evolve according to
the isothermal Euler equations in pressure--flux form
(Appendix~\ref{subsec:pq_transform}):
\begin{align}
\partial_t P_e + a^2 \, \partial_x q_e &= 0, \label{eq:setup_mass}\\
\partial_t q_e + \partial_x P_e + \partial_x\!\left(\frac{a^2 q_e^2}{P_e}\right)
+ \frac{f a^2 |q_e| q_e}{2 D P_e} &= 0, \label{eq:setup_mom}
\end{align}
for $x \in [0, L_e]$, $t \in [0, T]$, where $a$ is the isothermal
sound speed, $D$ the pipe diameter, and $f$ the Darcy friction factor.
\eqref{eq:setup_mass} expresses mass conservation;
\eqref{eq:setup_mom} balances inertia, pressure gradient, momentum
advection, and wall friction. Both nonlinear terms are singular as
$P \to 0$, so pressures remain strictly positive throughout.

\paragraph{Boundary and junction conditions.}
The vertex set splits as $\mc{V} = \mc{V}_D \,\dot\cup\, \mc{V}_J$ into boundary
vertices with prescribed pressure and internal junctions. $\mc{V}_D$
contains a single inlet $v_{\mathrm{in}}$ held at $P_{\mathrm{in}}$,
and outlet vertices held at a fixed $P_{\mathrm{out}} < P_{\mathrm{in}}$.
At each junction $v \in \mc{V}_J$, pressure is continuous and mass is
conserved:
\begin{align}
P_e(x_v, t) &= P_{e'}(x_v, t), && \forall\, e, e' \text{ incident to } v, \\
\sum_{e \in \mathrm{in}(v)} q_e(L_e, t) &= \sum_{e \in \mathrm{out}(v)} q_e(0, t).
\end{align}
Since all pipes share the same diameter, conservation of mass flow
reduces to conservation of mass flux.

\paragraph{Initial condition.}
The network starts at rest at the inlet pressure,
$P_e(x, 0) = P_{\mathrm{in}}$, $q_e(x, 0) = 0$ for all $e \in \mc{E}$.
This state is incompatible with the outlet condition: at $t = 0^+$ the
outlet pressure drops to $P_{\mathrm{out}}$, launching an expansion
wave into the network. The resulting transient, with reflections at
junctions and boundaries, relaxes toward a friction-dominated steady
state with nonzero throughflow. This transient is what the models must
capture.

\paragraph{Solution operator.}
With topology, pipe parameters, and $P_{\mathrm{out}}$ fixed, the
network state is determined by the inlet boundary data. We write the
solution operator as
\begin{equation}
S : [0, T] \times \mc{P} \to \mc{X},
\qquad (t, P_{\mathrm{in}}) \mapsto \big(P_e(\cdot, t), q_e(\cdot, t)\big)_{e \in \mc{E}},
\end{equation}
where $\mc{X}$ is the space of pressure–flux fields on $\mc{G}$ and
$\mc{P}$ is the space of admissible inlet data. In this work,
$\mc{P}$ is the one-parameter family of constant inlet pressures
$P_{\mathrm{in}} \in [P_{\mathrm{in}}^{\min}, P_{\mathrm{in}}^{\max}]$;
time-varying and multi-inlet boundary data are discussed in
Section~\ref{sec:limitations}.

\paragraph{Discrete target and learning task.}
Our target is the solution operator of the staggered finite-volume scheme of Appendix~\ref{sec:governing}, which carries the conservation structure and junction coupling the learned model inherits. The NBM
solver (Section~\ref{sec:method}) reproduces this scheme in the neural basis, and the learned model is trained on NBM reference trajectories at a finite set of inlet pressures in $[P_{\mathrm{in}}^{\min},
P_{\mathrm{in}}^{\max}]$. It returns the network state at any queried time in a single evaluation, without time stepping, for $P_{\mathrm{in}}$ inside that interval (interpolation) or beyond it (extrapolation). Errors are measured against these discrete references.

\section{Method Overview}
\label{sec:method}
\begin{figure}[h!]
\centering
\includegraphics[width=\textwidth]{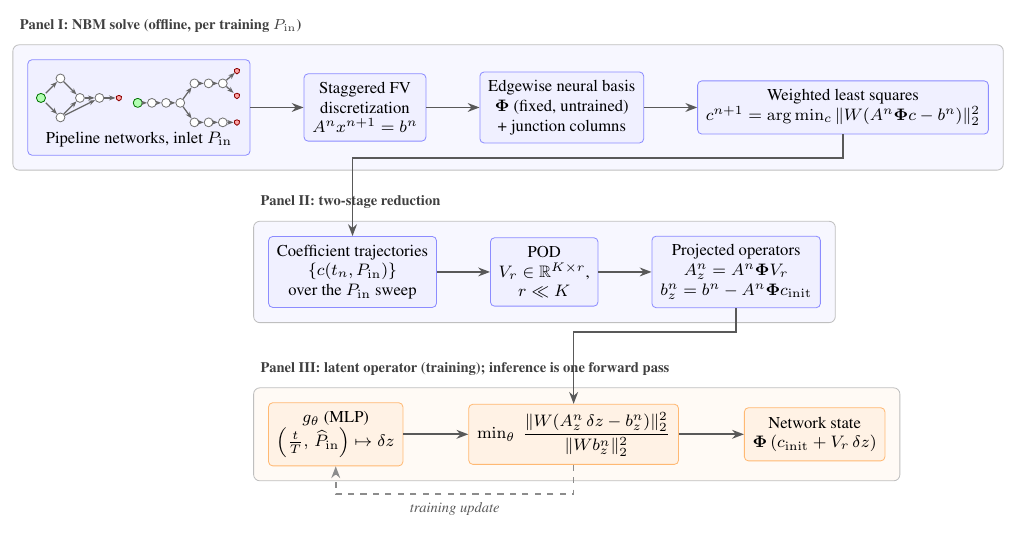}
\caption{Overview of NBM-OL. \textbf{Panel I}: for each training inlet
pressure, the staggered finite-volume system of the whole network,
including junction and boundary rows, is projected onto a fixed
edgewise neural basis and solved by weighted least squares, giving
coefficient trajectories; the thumbnails show the two topologies used
in our experiments (Section~\ref{sec:exp_setup}). \textbf{Panel II}:
POD compresses these trajectories to $r \ll K$ dimensions, and the
projected operators $(A_z^n, b_z^n)$ are precomputed.
\textbf{Panel III}: an MLP maps normalized time and inlet pressure to
the latent state, trained on the projected residual; at inference, one forward pass returns the network state at any
queried time.}
\label{fig:overview}
\end{figure}

NBM-OL proceeds in two stages, illustrated in
Figure~\ref{fig:overview}. The first is a deterministic solve: each
edge is equipped with a fixed neural basis
(Section~\ref{sec:basis}), and the staggered finite-volume equations
of the network, junction and boundary rows included, are projected
onto it and solved by weighted least squares at every time step
(Section~\ref{sec:projection}). Repeating this over a grid of inlet
pressures yields coefficient trajectories. The second stage learns in
a reduced coordinate: proper orthogonal decomposition compresses the
coefficients from $K$ to $r \ll K$ dimensions, and a network maps $(t, P_{\mathrm{in}})$ to
this latent state, trained on the projected discrete residual
(Section~\ref{sec:latent-op}). No basis function is ever trained; only
the latent map is. At inference, one forward pass returns the network
state at any queried time, with no time stepping.

\subsection{Neural Basis on a Network}
\label{sec:basis}

Following \citet{wang2026solving}, spatial bases are generated by a
fixed residual network with layers
\begin{equation}
h_\ell = \beta_\ell \, h_{\ell-1} P_\ell^\top
+ \tanh\!\big(\alpha_\ell \odot (h_{\ell-1} W_\ell^\top + b_\ell)\big),
\qquad \ell = 1, 2,
\end{equation}
where $W_\ell, P_\ell, b_\ell$ are fixed after random initialization,
$\alpha_\ell$ is a per-neuron shape vector, and $\beta_\ell$ a scalar
residual strength. The network is never trained. We evaluate it on the
normalized edge coordinate $x/L_e \in [0,1]$ at the finite-volume cell
centers (for $P$) and faces (for $q$), and orthonormalize the resulting columns, which controls the conditioning of the projected system. Since all edges share the same normalized discretization, one local basis per field, with $k$ functions, is reused on every edge.

The global basis $\Phi$ is block diagonal, with one pressure block and
one flux block per edge, and one additional unit column per internal
junction, so that each junction pressure is a coefficient. The
coefficient vector $c \in \mathbb{R}^K$ therefore has
$K = 2k|E| + |V_J|$ entries. Edges interact only through the junction
rows of the projected system below, which is what localizes NBM to
network domains: each edge carries its own expansion, coupled to its
neighbors through pressure continuity and mass conservation, and the
whole network is solved in a single monolithic projection.

\subsection{Projected Finite-Volume System}
\label{sec:projection}

The network equations are discretized by a staggered finite-volume
scheme with backward Euler time stepping and LUST reconstruction of
momentum advection (Appendix~\ref{sec:governing}). Freezing the
friction coefficient and advective velocity at the previous state
$u^n$ gives, at each step, a linear system
$A^n x^{n+1} = b^n$ over all cells, faces, junction, and boundary
rows, where both $A^n$ and $b^n$ depend on $u^n$. Substituting
$x^{n+1} = \Phi c^{n+1}$ yields an overdetermined system, solved as
\begin{equation}
c^{n+1} = \arg\min_{c} \big\| W \big(A^n \Phi c - b^n\big) \big\|_2^2,
\label{eq:nbm_step}
\end{equation}
where the diagonal matrix $W$ assigns a constant penalty weight to
boundary and junction rows and unit weight to interior rows. We solve
\eqref{eq:nbm_step} via the normal equations and a Cholesky
factorization. We use a constant penalty on boundary and junction rows rather than the energy-consistent weights of \citet{wang2026solving}, which places all residual rows on a common physical scale.

\subsection{Latent Operator and Training Objective}
\label{sec:latent-op}

A multilayer perceptron $g_\theta$ maps normalized time and inlet
pressure to the latent state,
\begin{equation}
\delta z(t, P_{\mathrm{in}}) = z_{\mathrm{scale}} \odot
g_\theta\!\left(\tfrac{t}{T},\,
\tfrac{P_{\mathrm{in}} - P_{\mathrm{in}}^{\min}}{P_{\mathrm{in}}^{\max} - P_{\mathrm{in}}^{\min}}\right),
\end{equation}
where $z_{\mathrm{scale}}$ is the componentwise standard deviation of
the training latents $V_r^\top(c_{\mathrm{ref}} - c_{\mathrm{init}})$,
offset by one. The predicted state is
$\Phi\big(c_{\mathrm{init}} + V_r\,\delta z\big)$: a combination of
fixed basis functions with coefficients computed from the inlet data,
i.e.\ the branch--trunk structure of DeepONet with a fixed trunk
$\Phi V_r$ derived from the discretized physics.

For a training pair $(t_{n+1}, P_{\mathrm{in}})$, we assemble $A^n$ and
$b^n$ of \eqref{eq:nbm_step} from the reference state
$u^n = \Phi\, c_{\mathrm{ref}}(t_n, P_{\mathrm{in}})$ and substitute
$c = c_{\mathrm{init}} + V_r \delta z$, giving the latent residual
\begin{equation}
R^n(\delta z) = W\big(A_z^n \delta z - b_z^n\big), \qquad
A_z^n = A^n \Phi V_r, \quad
b_z^n = b^n - A^n \Phi\, c_{\mathrm{init}}(P_{\mathrm{in}}).
\end{equation}
Training minimizes
\begin{equation}
\mc{L}(\theta) = \mathbb{E}\!\left[
\frac{1}{m}\frac{\| R^n(\delta z(t_{n+1}, P_{\mathrm{in}})) \|_2^2}{\| W b_z^n \|_2^2}
\right]
,
\label{eq:loss}
\end{equation}
over all training pairs, where $m$ is the number of rows. For each pair, the objective is minimized by the latent one-step solution from the reference state, so training fits reference next states in the
metric induced by the projected operator rather than in Euclidean coefficient distance. Reference data enter NBM-OL through the POD basis and the states $u^n$. No supervised loss term is used: the residual objective alone suffices. Because each $A_z^n$ has $r$ rather than $K$ columns, storing the projected operators for all $33{,}000$ training pairs of the 7-edge network requires $13.0$\,GB, whereas the full-coefficient equivalent would require $209.3$\,GB, a factor $K/r = 16.1$. Training hyperparameters are listed in Table~\ref{tab:nbmol_hparams}.

\paragraph{Residual as an error certificate.}
Fix a training pair, let $M = W A_z^n$ have full column rank, and let
$\delta z^\star$ minimize $\|R^n\|_2$. Since $R^n(\delta z^\star)$ is
orthogonal to the range of $M$,
\begin{equation}
\|\delta z - \delta z^\star\|_2 \le
\frac{\sqrt{\|R^n(\delta z)\|_2^2 - \|R^n(\delta z^\star)\|_2^2}}{\sigma_{\min}(M)}.
\label{eq:certificate}
\end{equation}
Since $\Phi$ has orthonormal columns and $V_r$ is orthonormal,
$\Phi V_r$ is an isometry, so the left-hand side equals the distance in grid values between the prediction and the best one-step solution representable in the latent space. The training objective therefore bounds a solution error, with a computable constant, rather than
serving only as an optimization surrogate.

\section{Experiments}
\label{sec:experiments}

We ask two questions: whether NBM reproduces the finite-volume scheme
on network domains (Section~\ref{sec:exp_verify}), and whether NBM-OL
learns the transient across inlet pressures, inside and outside the
training range (Section~\ref{sec:exp_ol}).

\subsection{Setup}
\label{sec:exp_setup}

\paragraph{Networks.}
We use two topologies with identical pipes ($L = 100$\,m, $D = 0.5$\,m,
$f = 0.015$, $a = 350$\,m/s) and discretization (200 cells per edge,
$\Delta t = 0.01$\,s, $T = 10$\,s); see Table~\ref{tab:fv_networks}.
The \emph{7-edge network} is a looped topology with two parallel paths
and a cross-link, one inlet, and one outlet. The \emph{tree network}
is a 12-edge branching topology adapted from a gas-network schematic
\citep{villalobos2003nlp}, with one inlet and three outlets over two
levels of branching; compressor stations in the schematic are modeled
as plain junctions. Outlets are held at $P_{\mathrm{out}} = 900$\,Pa.

\paragraph{Flow regime.}
Absolute pressures and pipe lengths are smaller than in transmission
pipelines; the dynamics are set by dimensionless ratios. A sound wave
crosses an edge $aT/L = 35$ times over the horizon, so the transient
is wave-dominated, with repeated reflections at junctions. At the highest inlet pressure tested, the steady-state Mach number reaches roughly $0.5$ on the most loaded edges, so momentum advection is not negligible and the problem is nonlinear throughout.

\paragraph{Metrics.}
We report relative $L^2$ errors, computed separately for pressure and
flux after concatenating all edges, spatial points (cell centers for
$P$, faces for $q$), and time steps into one vector per variable.

\subsection{Single-Instance Verification}
\label{sec:exp_verify}

For a fixed boundary condition ($P_{\mathrm{in}} = 1000$\,Pa), we
compare NBM against the finite-volume (FV) solution of the same
discretization, which NBM projects, and against a PINN
\citep{raissi2019physics} trained on the continuous PDE, boundary,
initial, and junction residuals without data (Appendix~\ref{app:pinn}).
The NBM--FV error measures how well the fixed basis reproduces the
scheme; the FV
discretization error is small compared with the PINN errors and does
not affect that comparison.

\begin{table}[h!]
\centering
\caption{Single-instance verification against the finite-volume
reference: relative $L^2$ error over all edges, spatial points, and
time steps.}
\label{tab:verify}
\begin{tabular}{lcccc}
\toprule
& \multicolumn{2}{c}{7-edge network} & \multicolumn{2}{c}{Tree network} \\
\cmidrule(lr){2-3} \cmidrule(lr){4-5}
Method & Pressure & Flux & Pressure & Flux \\
\midrule
NBM (ours) & $\mathbf{7.0\times10^{-6}}$ & $\mathbf{6.9\times10^{-5}}$
           & $\mathbf{1.0\times10^{-5}}$ & $\mathbf{1.3\times10^{-4}}$ \\
PINN & $2.0\times10^{-2}$ & $8.8\times10^{-1}$
     & $3.3\times10^{-2}$ & $9.2\times10^{-1}$ \\
\bottomrule
\end{tabular}
\end{table}

\begin{figure}[t]
\centering
\begin{subfigure}[b]{0.75\textwidth}
    \centering
    \includegraphics[width=\linewidth]{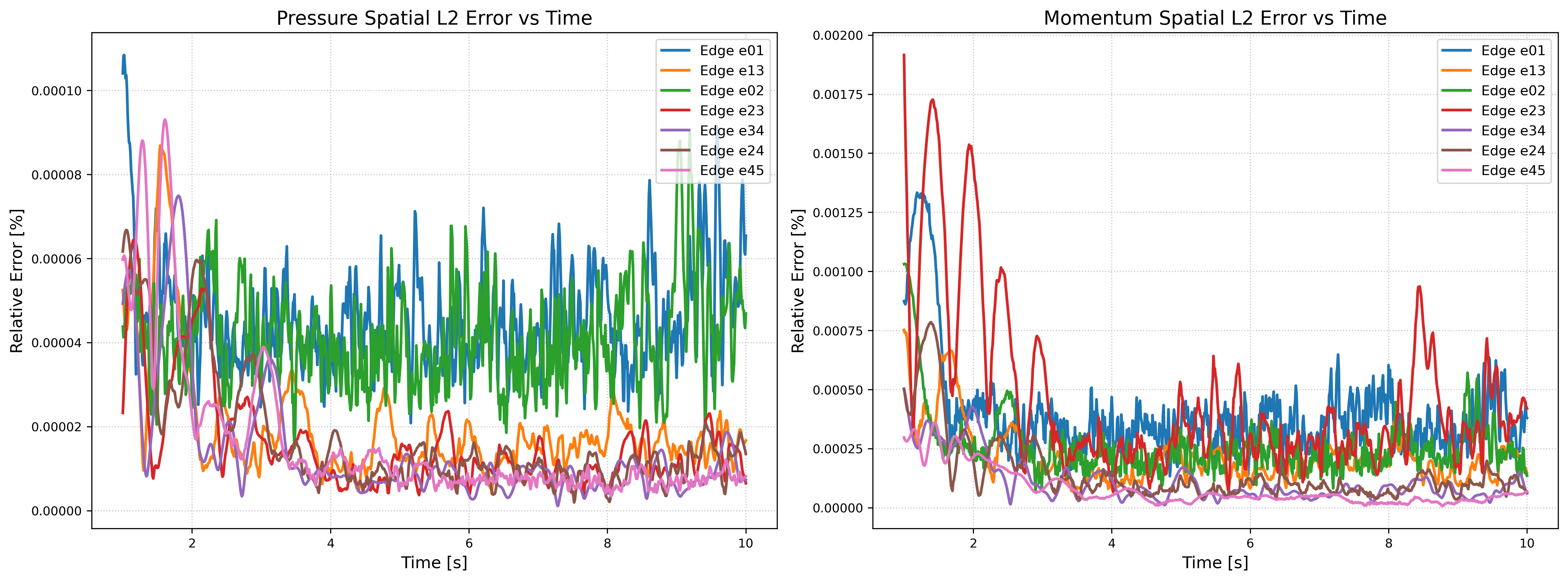}
    \caption{NBM vs.\ Finite Volume}
\end{subfigure}

\begin{subfigure}[b]{0.75\textwidth}
    \centering
    \includegraphics[width=\linewidth]{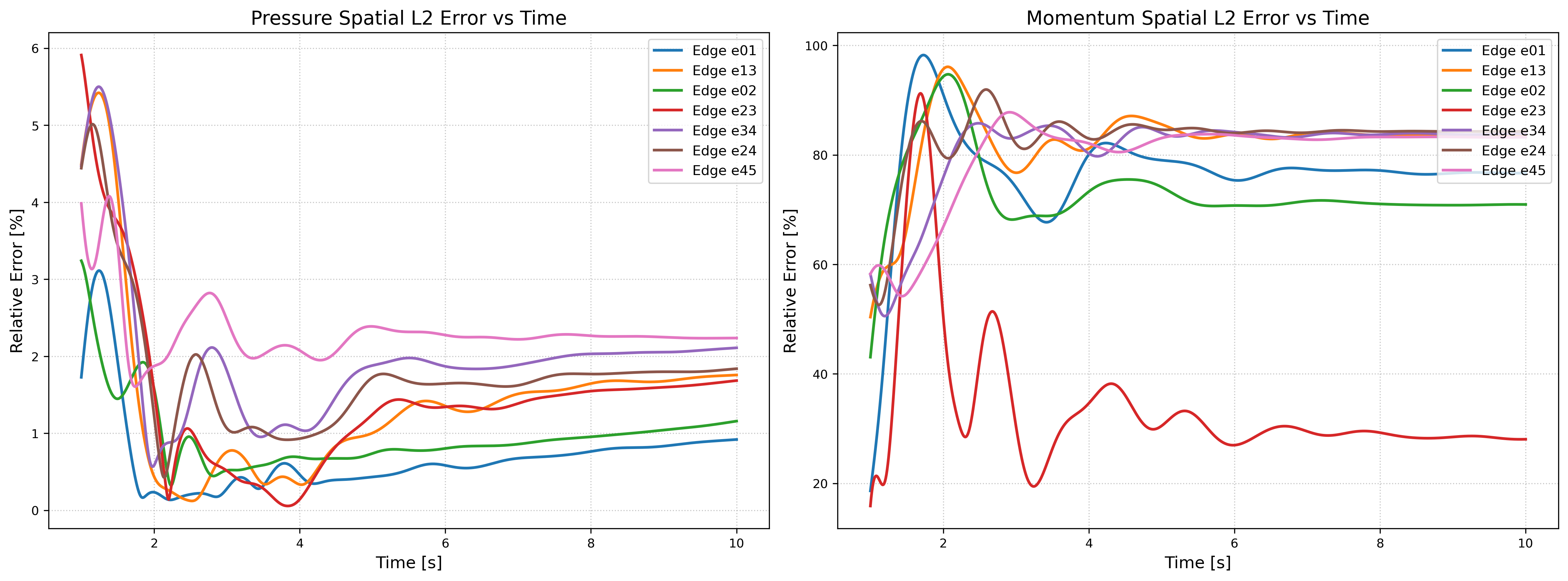}
    \caption{PINN vs.\ Finite Volume}
\end{subfigure}
\caption{Per-edge relative $L^2$ error over time, 7-edge network,
against the finite-volume reference: (a) NBM, (b) PINN. The
corresponding tree-network plots are in
Appendix~\ref{app:verify_tree}.}
\label{fig:7edge_l2-vs-time}
\end{figure}

NBM reproduces the scheme to relative errors of $10^{-5}$ in pressure
and $10^{-4}$ in flux on both networks (Table~\ref{tab:verify}), and
the per-edge error stays at this level over the whole transient
(Figure~\ref{fig:7edge_l2-vs-time}a). The PINN recovers pressure to a
few percent but not the flux, with errors of $88\%$ and $92\%$: it
does not capture the throughflow set up by the transient
(Figure~\ref{fig:7edge_l2-vs-time}b).

\subsection{Operator Learning}
\label{sec:exp_ol}

\paragraph{Setup.}
NBM-OL is trained on NBM trajectories at 33 inlet pressures on the
7-edge network and 20 on the tree network, evenly spaced in
$[1000, 1800]$\,Pa (Table~\ref{tab:nbmol_hparams}). We test at five
held-out interpolation points, none of them training pressures (on
the tree network, each lies midway between two training pressures),
and at two extrapolation points: $1840$ and $1880$\,Pa on the 7-edge
network, and $1820$ and $1880$\,Pa on the tree network, i.e.\ $2.5$--$10\%$
of the training range beyond its upper end. Errors are measured against the NBM reference, which differs from FV
by less than $0.02\%$ (Table~\ref{tab:verify}).

\paragraph{Baseline.}
PI-DeepONet \citep{wang2021learning}, adapted to networks with an edge
one-hot input to the trunk, is trained on the same inlet-pressure range
from PDE, boundary, initial, and junction residuals
(Appendix~\ref{app:pideeponet}).

\begin{table}[H]
\centering
\caption{Operator learning: relative $L^2$ error (\%) averaged over
held-out inlet pressures. Per-point errors are in
Appendix~\ref{app:ol_perpoint}.}
\label{tab:ol_main}
\begin{tabular}{llcccc}
\toprule
& & \multicolumn{2}{c}{Interpolation} & \multicolumn{2}{c}{Extrapolation} \\
\cmidrule(lr){3-4} \cmidrule(lr){5-6}
Network & Method & Pressure & Flux & Pressure & Flux \\
\midrule
\multirow{2}{*}{7-edge} & NBM-OL (ours) & $\mathbf{0.065}$ & $\mathbf{1.05}$ & $\mathbf{0.23}$ & $\mathbf{1.62}$ \\
                        & PI-DeepONet   & $7.90$ & $87.3$ & $11.8$ & $85.4$ \\
\midrule
\multirow{2}{*}{Tree}   & NBM-OL (ours) & $\mathbf{0.12}$ & $\mathbf{3.25}$ & $\mathbf{0.97}$ & $\mathbf{2.84}$ \\
                        & PI-DeepONet   & $13.0$ & $95.3$ & $17.9$ & $94.1$ \\
\bottomrule
\end{tabular}
\end{table}

\begin{figure}[H]
\centering
\begin{subfigure}[b]{\linewidth}
    \includegraphics[width=\linewidth]{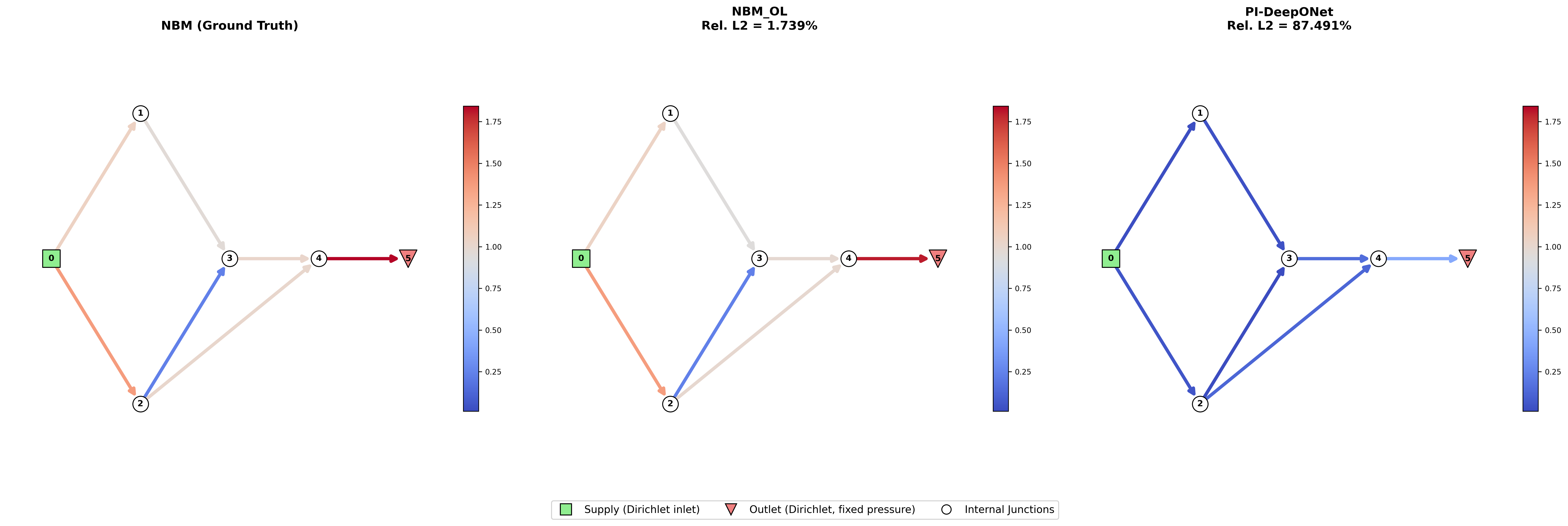}
    \caption{Network heatmap, momentum $q$ at $t=2.0$s}
\end{subfigure}
\vspace{4pt}
\begin{subfigure}[b]{\linewidth}
    \includegraphics[width=\linewidth]{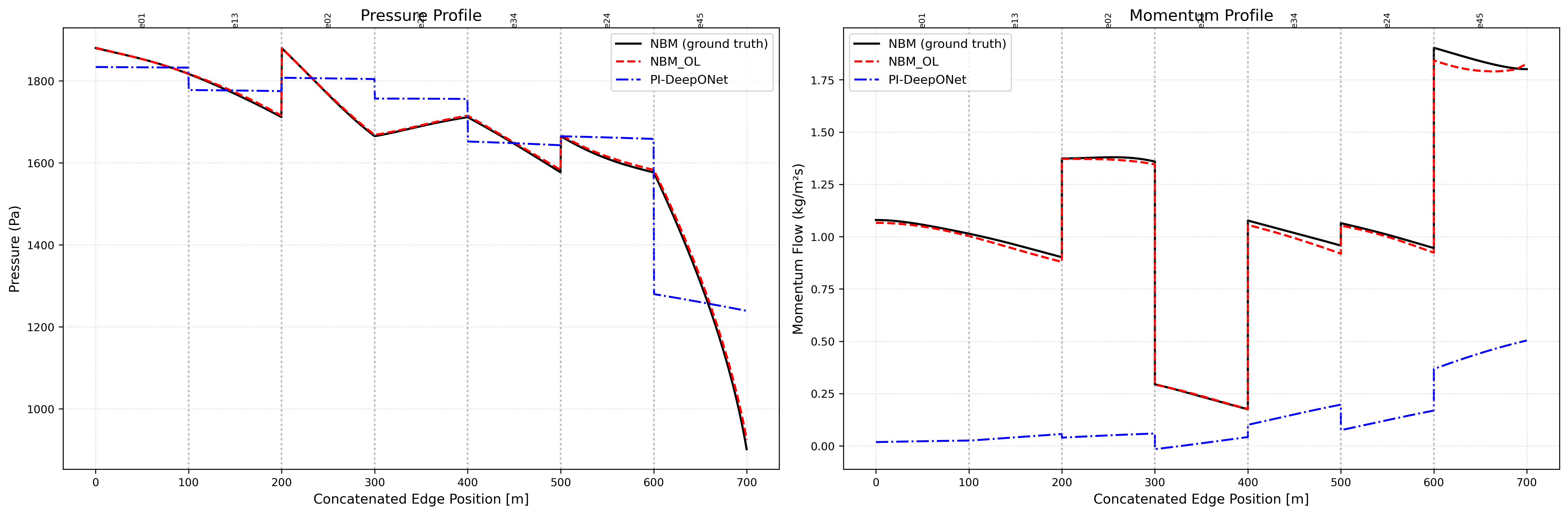}
    \caption{Spatial profile at $t=2.0$s}
\end{subfigure}
\vspace{4pt}
\begin{subfigure}[b]{\linewidth}
    \includegraphics[width=\linewidth]{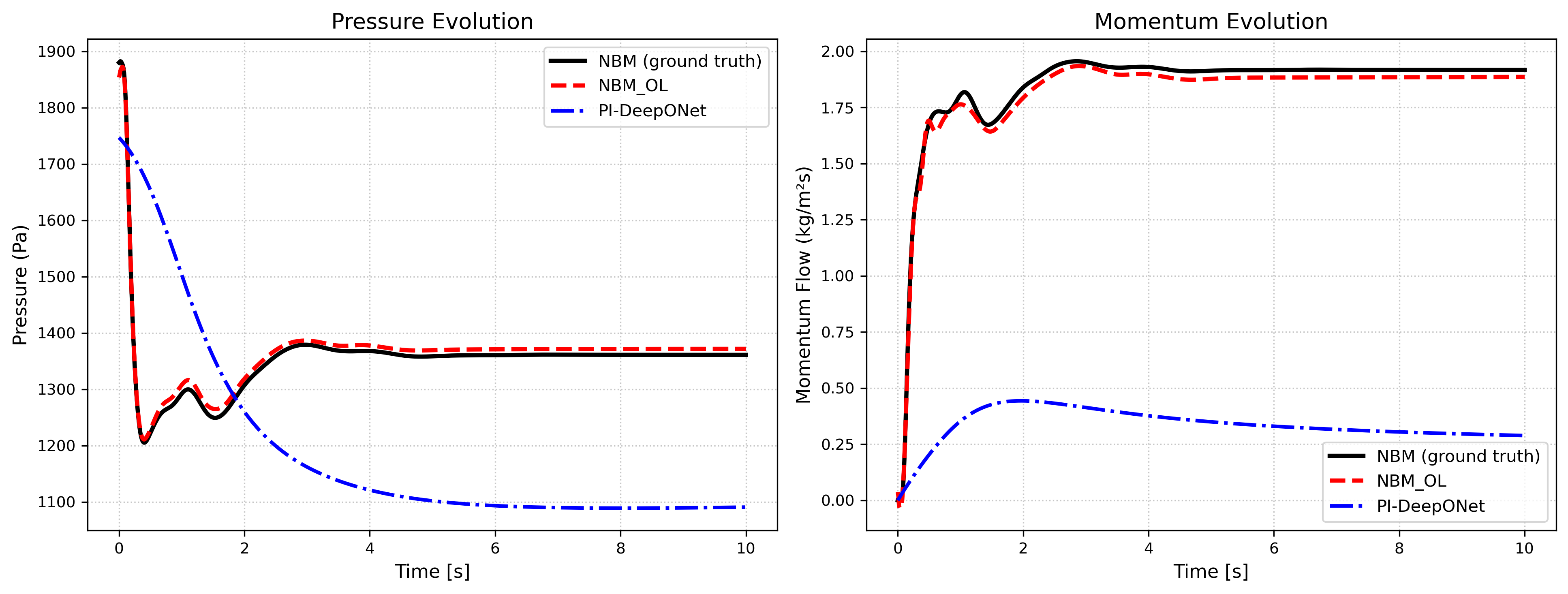}
    \caption{Time evolution, edge e45}
\end{subfigure}
\caption{Extrapolation, 7-edge network, $P_{\mathrm{in}} = 1880$\,Pa
($10\%$ of the training range beyond its upper end). Reference NBM
(black), NBM-OL (red), PI-DeepONet (blue). (a) Flux over the network
at $t = 2$\,s. (b) Pressure and flux profiles along the concatenated
edges at $t = 2$\,s. (c) Time evolution at the midpoint of edge e45. Extrapolation, tree network, $P_{\mathrm{in}} = 1880$\,Pa is in Appendix~\ref{app:tree_extrap}.}
\label{fig:extrap_7edge_combined}
\end{figure}

\paragraph{Results.}
NBM-OL's mean errors are $18$--$122\times$ lower than PI-DeepONet's in
pressure and $29$--$83\times$ lower in flux across both networks and
both regimes (Table~\ref{tab:ol_main}). In interpolation, pressure
errors stay at or below $0.15\%$ at every test point. Flux errors are
largest at the lowest test pressures ($1.32\%$ at $1215$\,Pa on the
7-edge network, $7.15\%$ at $1063.2$\,Pa on the tree network) and fall
as $P_{\mathrm{in}}$ increases. This is consistent with the driving
pressure difference $P_{\mathrm{in}} - P_{\mathrm{out}}$, and hence the
throughflow, being smallest there, so that a similar absolute error
yields a larger relative one.

Under extrapolation, NBM-OL pressure error grows with distance from the
training range, most visibly on the tree network, from $0.24\%$ at
$+2.5\%$ to $1.69\%$ at $+10\%$. Flux error at $+10\%$ is at most about
twice the interpolation mean. Figures~\ref{fig:extrap_7edge_combined}
and~\ref{fig:extrap_tree_combined} show the $+10\%$ case: NBM-OL
follows the transient and the steady state closely, with a small
steady-state offset on some edges.

PI-DeepONet reaches flux errors of $85$--$96\%$ at every test point,
in both regimes. Its predicted throughflow remains a small fraction of
the true value (Figures~\ref{fig:extrap_7edge_combined}
and~\ref{fig:extrap_tree_combined}), and its pressure profiles are
distorted at junctions. As with the PINN, the physics-only objective
fails to establish the throughflow, the feature of the solution that
the network coupling determines.

\section{Conclusions and Future Work}
\label{sec:limitations}

We developed the Neural Basis Method to fluid flow networks. Edgewise
neural bases, coupled through junction conditions in a single
monolithic projection of a staggered finite-volume scheme, reproduce
the discrete solution of the isothermal Euler equations to relative
errors of $10^{-5}$ in pressure and $10^{-4}$ in flux on looped and
branching networks. Building on this representation, NBM-OL maps time
and inlet pressure to a POD-compressed latent state, trained with the
projected discrete residual along reference trajectories. At held-out inlet pressures, inside the training
range and up to $10\%$ of its width beyond it, mean errors stay below
$1\%$ in pressure and $3.3\%$ in flux, one to two orders of magnitude
below a physics-informed DeepONet, and a single forward pass returns
the network state at any queried time.

Several directions point to future work. \emph{Boundary data}: the
learned operator is exercised on a one-parameter family of constant
inlet pressures; time-varying and multi-inlet boundary data change
only the input to the coefficient network, but remain to be tested.
\emph{Scale and regime}: the construction is local by design, since
each edge carries its own basis and edges meet only through junction
rows, so larger networks enlarge the projected system without changing its structure; applying it at transmission scale, with kilometer-long pipes and megapascal pressures, is a direct next step.
\emph{Extrapolation}: the latent basis is built from training
trajectories, so predictions outside the training range are limited
by its span; our tests extend only $10\%$ beyond it, and pressure error grows with distance. \emph{Reference data}: NBM-OL uses reference coefficient trajectories
for its POD basis and for the states entering the training residual. Inference requires neither: a single forward pass at $(t, P_{\mathrm{in}})$ returns the state. Assembling the residual from the model's own predictions would remove the remaining dependence and make the
residual computable as a self-check at inference as well.

In our experiments, placing the physics in a fixed, network-aware
discretization rather than in a penalty loss yields a surrogate that
follows the transient and establishes the throughflow, which the
physics-only neural baselines did not. Extending this construction to
richer boundary data and larger networks is a step toward fast,
reliable surrogates for many-query operation of fluid networks.

\section*{Reproducibility Statement}

Section~\ref{sec:method} specifies the NBM-OL construction: the neural
basis and its junction coupling, the weighted least-squares projection,
the POD reduction, and the training objective.
Appendix~\ref{sec:governing} gives the governing equations and the
staggered finite-volume discretization, including the LUST
reconstruction, boundary treatment, and lagged-coefficient time
integration; Appendix~\ref{sec:nbm} gives the basis construction and
the projected coefficient-space system. Network configurations for
both topologies are in Table~\ref{tab:fv_networks}, and all NBM-OL
training settings -- training grid, latent dimension, architecture,
optimizer, and schedule -- in
Table~\ref{tab:nbmol_hparams}. Section~\ref{sec:experiments} states
the evaluation protocol, including the held-out $P_{\text{in}}$ values
for both topologies, which are listed individually in
Appendix~\ref{app:ol_perpoint}. Appendices~\ref{app:pinn}
and~\ref{app:pideeponet} give the architecture, loss formulation
(including the nondimensionalization of the PDE residual,
\eqref{eq:q_ref}), collocation-point counts
(Table~\ref{tab:pideeponet_colloc}), and training hyperparameters
(Tables~\ref{tab:pinn_hyperparams} and~\ref{tab:pideeponet_hparams})
for the two baselines, applied identically across both topologies
except where topology-derived quantities necessarily differ. All runs
use a fixed random seed. Code for both network topologies, together with trained model
checkpoints, will be made publicly available upon publication.

\subsection*{AI use statement}
The research idea, methodology, experimental design, and findings are
the authors' own. Generative AI tools were used to assist with writing
and editing code, formatting tables, and editing parts of the
manuscript for readability. All AI-assisted code was executed and its
output verified by the authors, who take full responsibility for the
content of this manuscript.

\bibliographystyle{plainnat}
\bibliography{main}

\appendix
\section*{Appendix}

\section{Governing Equations and FV Discretization}
\label{sec:governing}

\subsection{Conservative Formulation}
\label{subsec:conservative}

The dynamics of gas flow in a single pipe are governed by the one-dimensional
isothermal Euler equations. We write these in conservative form, using fluid
density $\rho$ and mass flux $m$ as the primary variables:
\begin{align}
    \frac{\partial \rho}{\partial t} + \frac{\partial m}{\partial x} &= 0,
    \label{eq:mass_conservative} \\[4pt]
    \frac{\partial m}{\partial t} + \frac{\partial}{\partial x}\!\left(\frac{m^2}{\rho}\right)
    + a^2 \frac{\partial \rho}{\partial x} + \frac{f |m| m}{2 D \rho} &= 0.
    \label{eq:momentum_conservative}
\end{align}
Here $a$ is the isothermal speed of sound, $D$ is the pipe diameter, and $f$
is the Darcy friction factor. The pressure gradient term in
\eqref{eq:momentum_conservative} already reflects the isothermal equation
of state $P = \rho a^2$.

\subsection{Transformation to Pressure--Flow Variables}
\label{subsec:pq_transform}

Density is not the most convenient variable for network assembly, since
pressure is the quantity naturally shared across pipe junctions. We
eliminate $\rho$ in favor of $P$ using the isothermal equation of state:
\begin{equation}
    \rho = \frac{P}{a^2}
    \quad\Longrightarrow\quad
    \frac{\partial \rho}{\partial t} = \frac{1}{a^2}\frac{\partial P}{\partial t},
    \qquad
    \frac{\partial \rho}{\partial x} = \frac{1}{a^2}\frac{\partial P}{\partial x}.
    \label{eq:eos_substitution}
\end{equation}
Substituting \eqref{eq:eos_substitution} into
\eqref{eq:mass_conservative}--\eqref{eq:momentum_conservative} casts the
system entirely in $(P,q)$ variable space, where $q \equiv m = \rho u$ is the mass flux per unit cross-sectional area; the same quantity is the momentum per unit volume, which is why it is the natural unknown in \eqref{eq:momentum_pq}:
\begin{align}
    \frac{\partial P}{\partial t} + a^2 \frac{\partial q}{\partial x} &= 0,
    \label{eq:mass_pq} \\[4pt]
    \frac{\partial q}{\partial t} + \frac{\partial P}{\partial x}
    + \frac{\partial}{\partial x}\!\left(\frac{a^2 q^2}{P}\right)
    + \frac{f a^2 |q| q}{2 D P} &= 0.
    \label{eq:momentum_pq}
\end{align}
We work with this formulation throughout the paper. Both
the advective and friction terms in \eqref{eq:momentum_pq} are singular
as $P \to 0$; strictly positive boundary and initial pressures are therefore
required to keep the assembled system well-conditioned.

\subsection{Network Formulation}
\label{subsec:network}

The pipeline network is represented as a directed metric graph
$\mc{G} = (\mc{V}, \mc{E})$, with each edge
$e \in \mc{E}$ an interval $[0, L_e]$ on which
\eqref{eq:mass_pq}--\eqref{eq:momentum_pq} hold independently. At
every internal (non-boundary) vertex $k \in \mathcal{V}$, two coupling
conditions are enforced: continuity of pressure across all incident edges,
and mass conservation,
\begin{equation}
    \sum_{e:\, e \to k} q_e(L_e, t) \;=\; \sum_{e:\, k \to e} q_e(0, t),
    \label{eq:kirchhoff}
\end{equation}
i.e., total incoming momentum flow equals total outgoing momentum flow,
with no accumulation at internal vertices. Boundary vertices are held at a
prescribed Dirichlet pressure.

\subsection{Spatial Discretization}
\label{subsec:strategy}

We solve \eqref{eq:mass_pq}--\eqref{eq:momentum_pq} using a
finite-volume method on a staggered mesh. Pressure is stored at cell
centers, indexed by $i$. Momentum is stored at cell faces, indexed by $j$,
so that the two boundary faces of an edge coincide with its incident
vertices. This choice avoids pressure--velocity decoupling, a well-known
failure mode on collocated grids that manifests as checkerboard
oscillations in the pressure field.

\subsection{Discrete Mass Equation}
\label{subsec:discrete_mass}

We integrate the mass equation, \eqref{eq:mass_pq}, over the finite
volume centered at pressure node $P_i$. Its boundary faces coincide with
momentum nodes $q_j$ and $q_{j+1}$. Applying the divergence theorem gives
\begin{equation}
    \frac{P_i^{n+1} - P_i^n}{\Delta t}
    + a^2 \frac{q_{j+1}^{n+1} - q_j^{n+1}}{\Delta x} = 0.
    \label{eq:mass_discrete_raw}
\end{equation}
We collect all $t^{n+1}$ unknowns on the left-hand side. This yields the
row contributed by node $i$ to the global linear system
$\mathbf{A}\mathbf{x} = \mathbf{b}$:
\begin{equation}
    \left(\frac{1}{\Delta t}\right) P_i^{n+1}
    + \left(\frac{a^2}{\Delta x}\right) q_{j+1}^{n+1}
    - \left(\frac{a^2}{\Delta x}\right) q_j^{n+1}
    = \frac{P_i^n}{\Delta t}.
    \label{eq:mass_discrete_assembled}
\end{equation}

\subsection{Discrete Momentum Equation}
\label{subsec:discrete_momentum}

We integrate the momentum equation, \eqref{eq:momentum_pq}, over the
control volume centered at momentum face $q_j$. This gives
\begin{equation}
    \frac{q_j^{n+1} - q_j^n}{\Delta t}
    + \frac{P_i^{n+1} - P_{i-1}^{n+1}}{\Delta x}
    + C_{\text{fric}}^n\, q_j^{n+1}
    + \frac{F_{\text{LUST},i}^{n+1} - F_{\text{LUST},i-1}^{n+1}}{\Delta x} = 0,
    \label{eq:momentum_discrete_raw}
\end{equation}
where $F_{\text{LUST}}$ is the advective momentum flux, reconstructed as
described in Section~\ref{subsec:lust}. Collecting $t^{n+1}$ unknowns gives
\begin{equation}
    \left(\frac{1}{\Delta t} + C_{\text{fric}}^n\right) q_j^{n+1}
    + \left(\frac{1}{\Delta x}\right) P_i^{n+1}
    - \left(\frac{1}{\Delta x}\right) P_{i-1}^{n+1}
    + \big[\text{LUST stencil}\big]
    = \frac{q_j^n}{\Delta t}.
    \label{eq:momentum_discrete_assembled}
\end{equation}

\subsection{LUST Reconstruction}
\label{subsec:lust}

The flux term $F_{\text{LUST}}$ requires momentum at pressure-cell
centers, where no momentum degree of freedom lives; we reconstruct it
from face values using a Linear Upwind Stabilised Transport (LUST)
scheme \citep{weller2012controlling}, a blend of central differencing
(second-order accurate but dispersive near sharp gradients) and
second-order linear upwind differencing, which introduces numerical
dissipation \citep{rezaeiravesh2019systematic}. For a lagged advective
velocity $u_{\text{adv}}^n \geq 0$, the reconstructed center value is
\begin{equation}
    q_{\text{center},i}^{n+1}
    = \beta \underbrace{\left(\frac{q_j^{n+1} + q_{j+1}^{n+1}}{2}\right)}_{\text{CD}}
    + (1-\beta) \underbrace{\left(1.5\, q_j^{n+1} - 0.5\, q_{j-1}^{n+1}\right)}_{\text{LUD}},
    \label{eq:lust_reconstruction}
\end{equation}
with the stencil reversed for $u_{\text{adv}}^n < 0$ and a reduced
two-point stencil at boundary-adjacent cells. We use $\beta = 0.75$,
the standard LUST weighting \citep{rezaeiravesh2019systematic}. The
reconstructed flux enters the momentum equation as
\begin{equation}
    F_{\text{LUST},i}^{n+1} = u_{\text{adv}}^n \cdot q_{\text{center},i}^{n+1}.
    \label{eq:lust_flux}
\end{equation}

\subsection{Boundary Treatment}
\label{subsec:boundary}

At each edge endpoint, the momentum equation is discretized over a
half-cell (of width $\Delta x / 2$) between the boundary face and the
first interior cell center, giving a boundary-face pressure-gradient term
scaled by $2/\Delta x$ rather than $1/\Delta x$. For an edge endpoint
incident to a Dirichlet (boundary) vertex, the corresponding node pressure
is substituted as a known constant; for an endpoint incident to an
internal vertex, the vertex's own (unknown) pressure is included in the
linear system alongside the Kirchhoff constraint (Eq.~\ref{eq:kirchhoff}).

\subsection{Time Integration}
\label{subsec:temporal}

Spatial discretization of Section~\ref{sec:governing} reduces
\eqref{eq:mass_pq}--\eqref{eq:momentum_pq} to a semi-discrete system
$d\mathbf{U}/dt = \mathbf{F}(\mathbf{U})$, where $\mathbf{U}$ collects all
pressure and momentum degrees of freedom. We advance this system with
Backward Euler,
\begin{equation}
    \mathbf{U}^{n+1} = \mathbf{U}^n + \Delta t\, \mathbf{F}(\mathbf{U}^{n+1}).
    \label{eq:backward_euler}
\end{equation}

$\mathbf{F}$ is nonlinear: friction depends on $|q|$, and advection on a
velocity computed from the solution itself. Solving
\eqref{eq:backward_euler} exactly would require an inner Newton
iteration at every step. We avoid this by linearizing $\mathbf{F}$ around
the previous state, freezing its nonlinear coefficients at $t^n$ so that
\eqref{eq:backward_euler} reduces to a linear system in
$\mathbf{U}^{n+1}$,
\begin{equation}
    \big(\mathbf{I} - \Delta t\, \mathbf{A}^n\big)\, \mathbf{U}^{n+1}
    = \mathbf{U}^n + \Delta t\, \mathbf{b}^n,
    \label{eq:be_linear_system}
\end{equation}
where $\mathbf{A}^n$ and $\mathbf{b}^n$ depend only on known, $t^n$-level
quantities.

Two coefficients enter $\mathbf{A}^n$. The friction coefficient,
\begin{equation}
    C_{\text{fric}}^n = \frac{f a^2 |q^n|}{2 D \bar{P}_{\text{face}}^n},
    \label{eq:fric_coeff}
\end{equation}
uses $\bar{P}_{\text{face}}^n$, the arithmetic average of the two
pressure cells adjacent to the face. The advective velocity,
\begin{equation}
    u_{\text{adv}}^n = \frac{a^2 \bar{q}_{\text{center}}^n}{P_{\text{center}}^n},
    \label{eq:adv_velocity}
\end{equation}
uses $\bar{q}_{\text{center}}^n$, a cell-centered momentum reconstructed
from the two adjacent faces.

\section{The Neural Basis Method}
\label{sec:nbm}

\subsection{Basis Representation}
\label{subsec:nbm_basis}

The discretization described in Section~\ref{sec:governing} produces one
unknown per physical grid location: one pressure value per cell, one
momentum value per face. For a network of any practical size, this yields a
prohibitively large state vector. The Neural Basis Method reduces this
dimensionality by representing each field as a linear combination of a
small number of fixed spatial basis functions, with only the combination
coefficients evolving in time:
\begin{equation}
    P^{n+1}(x) \approx \boldsymbol{\Phi}_P(x)\, \mathbf{c}_P^{n+1},
    \qquad
    q^{n+1}(x) \approx \boldsymbol{\Phi}_q(x)\, \mathbf{c}_q^{n+1}.
    \label{eq:nbm_basis_rep}
\end{equation}
Here $\boldsymbol{\Phi}_P$ and $\boldsymbol{\Phi}_q$ are fixed, precomputed
basis matrices evaluated at the pressure cell centers and momentum faces
respectively (constructed as described in
Section~\ref{subsec:nbm_construction}), and $\mathbf{c}_P^{n+1}$,
$\mathbf{c}_q^{n+1}$ are the time-dependent coefficient vectors we solve
for at each step, in place of the full physical state. The number of
coefficients is chosen to be much smaller than the number of physical
degrees of freedom; the resulting compression is what makes network-scale
simulation tractable.

\subsection{Basis Construction}
\label{subsec:nbm_construction}

For each field (pressure and momentum), a shared local
basis is obtained by evaluating a fixed, randomly initialized neural
network at the field's grid locations (cell centers for pressure, faces
for momentum) on the edge's normalized domain $[0,1]$, then
orthonormalizing the resulting evaluations via QR decomposition.

Because every edge shares the same length, spatial resolution, and
non-dimensionalized domain, this local basis is computed once and reused
across all edges, placed in a block-diagonal arrangement within the global
basis matrix $\boldsymbol{\Phi}$: edge $e$'s pressure coefficients occupy
their own dedicated block of columns of $\boldsymbol{\Phi}$, multiplying
only that edge's pressure degrees of freedom, and likewise for momentum.
Internal junction vertices are not represented by a spatial basis
expansion (a single point has no spatial extent to expand), and instead
receive one dedicated coefficient each, entering $\boldsymbol{\Phi}$ as a
single unit entry in its own column, so that an internal vertex's
coefficient equals its pressure directly.

\subsection{Projected Discrete Equations}
\label{subsec:nbm_projection}

We obtain the coefficient-space equations by substituting the basis
representation, \eqref{eq:nbm_basis_rep}, directly into the assembled
finite volume equations of Section~\ref{sec:governing}. This substitution
is performed independently at every cell center $i$ and every momentum
face $j$.

Substituting into the discrete mass equation,
\eqref{eq:mass_discrete_assembled}, at center $i$ gives
\begin{equation}
    \left[\frac{1}{\Delta t}\, \boldsymbol{\Phi}_P(x_i)\right] \mathbf{c}_P^{n+1}
    + \left[\frac{a^2}{\Delta x}\,
        \big(\boldsymbol{\Phi}_q(x_{j+1}) - \boldsymbol{\Phi}_q(x_j)\big)\right] \mathbf{c}_q^{n+1}
    = \frac{P_i^n}{\Delta t}.
    \label{eq:nbm_mass_row}
\end{equation}
Substituting into the discrete momentum equation,
\eqref{eq:momentum_discrete_assembled}, at face $j$ gives
\begin{equation}
    \left[\frac{\boldsymbol{\Phi}_P(x_i) - \boldsymbol{\Phi}_P(x_{i-1})}{\Delta x}\right] \mathbf{c}_P^{n+1}
    + \left[\left(\frac{1}{\Delta t} + C_{\text{fric}}^n\right)
        \boldsymbol{\Phi}_q(x_j)\right] \mathbf{c}_q^{n+1}
    = \frac{q_j^n}{\Delta t}.
    \label{eq:nbm_momentum_row}
\end{equation}
Junction rows (Eq.~\ref{eq:kirchhoff}) and boundary rows
(Section~\ref{subsec:boundary}) are projected identically, substituting
$\boldsymbol{\Phi}_q$ (and, at an internal vertex, that vertex's own unit
basis column) in place of the raw physical unknowns.

\subsection{Global Coefficient-Space System}
\label{subsec:nbm_global_system}

Collecting the coefficients into a single unknown vector $\mathbf{x} =
[\mathbf{c}_P^{n+1};\, \mathbf{c}_q^{n+1}]^\top$, the assembled rows from
\eqref{eq:nbm_mass_row} and \eqref{eq:nbm_momentum_row}, taken over the
entire network, form the global block system
\begin{equation}
    \begin{bmatrix}
        \mathbf{A}_{PP} & \mathbf{A}_{Pq} \\
        \mathbf{A}_{qP} & \mathbf{A}_{qq}
    \end{bmatrix}
    \begin{bmatrix}
        \mathbf{c}_P^{n+1} \\
        \mathbf{c}_q^{n+1}
    \end{bmatrix}
    =
    \begin{bmatrix}
        \mathbf{b}_P \\
        \mathbf{b}_q
    \end{bmatrix},
    \label{eq:nbm_global_system}
\end{equation}
with block entries defined row-wise by
\begin{align}
    \mathbf{A}_{PP} &= \frac{1}{\Delta t}\, \boldsymbol{\Phi}_P(x_i),
    \label{eq:app} \\
    \mathbf{A}_{Pq} &= \frac{a^2}{\Delta x}\,
        \big(\boldsymbol{\Phi}_q(x_{j+1}) - \boldsymbol{\Phi}_q(x_j)\big),
    \label{eq:apq} \\
    \mathbf{A}_{qP} &= \frac{1}{\Delta x}\,
        \big(\boldsymbol{\Phi}_P(x_i) - \boldsymbol{\Phi}_P(x_{i-1})\big),
    \label{eq:aqp} \\
    \mathbf{A}_{qq} &= \frac{1}{\Delta t}\, \boldsymbol{\Phi}_q(x_j)
        + C_{\text{fric}}^n\, \boldsymbol{\Phi}_q(x_j) + \text{[LUST terms]}.
    \label{eq:aqq}
\end{align}
The bracketed LUST terms in \eqref{eq:aqq} are the projection of the
advective stencil: substituting \eqref{eq:lust_reconstruction} into
\eqref{eq:lust_flux} and then into the flux difference of
\eqref{eq:momentum_discrete_raw} gives a linear combination of
$\boldsymbol{\Phi}_q$ evaluated at the faces neighbouring $x_j$, with
coefficients set by $\beta$, $u_{\text{adv}}^n$, and the upwind
direction.

Because \eqref{eq:nbm_global_system} is assembled from one row per
physical cell and face, but only a small number of coefficients are being
solved for, the resulting system is rectangular and overdetermined: the
number of rows equals the number of physical degrees of freedom in the
original finite volume discretization, while the number of columns equals
the number of basis coefficients, chosen to be much smaller. No exact
solution exists in general; we instead solve
\eqref{eq:nbm_global_system} in the weighted least-squares sense,
\begin{equation}
    \mathbf{x}^{n+1} = \operatorname*{arg\,min}_{\mathbf{x}}
    \left\| \mathbf{W} \big(\mathbf{A}\mathbf{x} - \mathbf{b}\big) \right\|_2^2,
    \label{eq:nbm_wls}
\end{equation}
where $\mathbf{W}$ is a diagonal weighting matrix, with a substantially
larger weight assigned to boundary and junction rows than to interior
rows. This allows the solver to prioritize satisfying boundary conditions
and junction mass conservation over the interior rows of
\eqref{eq:nbm_global_system}.

\subsection{Network Instantiations}
\label{subsec:network_instantiations}

Both the finite-volume reference solver and the Neural Basis Method are
applied, without modification to the governing equations or numerical
scheme, to two network topologies differing only in graph structure and
boundary configuration, summarized in Table~\ref{tab:fv_networks}.

\begin{table}[h]
\centering
\caption{Network configurations for the two topologies. Both use
identical pipe parameters ($a = 350$ m/s, $D = 0.5$ m, $f = 0.015$) and
per-edge resolution ($L = 100$ m, $\Delta x = 0.5$ m, $N = 200$
cells/edge).}
\label{tab:fv_networks}
\begin{tabular}{lcc}
\toprule
& 7-edge network & Tree network \\
\midrule
Edges $|\mathcal{E}|$ & 7 & 12 \\
Vertices $|\mathcal{V}|$ & 6 & 13 \\
Topology & 2 parallel paths, 1 cross-link & Single-inlet branching tree \\
Dirichlet inlets & 1 (node 0) & 1 (node 0) \\
Dirichlet outlets & 1 (node 5) & 3 (nodes 7, 8, 12) \\
Internal (junction) vertices & 4 & 9 \\
$\Delta t$ & 0.01 s & 0.01 s \\
$T_{\text{final}}$ & 10 s & 10 s \\
\bottomrule
\end{tabular}
\end{table}

\begin{table}[h]
\centering
\caption{NBM solver configuration.}
\label{tab:nbm_config}
\begin{tabular}{lcc}
\toprule
 & 7-edge & Tree \\
\midrule
Basis functions per field per edge & 40 & 40 \\
Basis network & \multicolumn{2}{c}{2-layer residual net, widths [128, 40], $\tanh$}  \\
Boundary/junction penalty weight & $10^4$ & $10^4$ \\
Grid degrees of freedom & 2811 & 4821 \\
NBM coefficients $K$ & 564 & 969 \\
\bottomrule
\end{tabular}
\end{table}

\section{NBM-OL Implementation}
\label{app:nbmol}

\paragraph{Data generation.}
For each network we run the NBM solver
(Section~\ref{sec:method}) once per training inlet pressure on a
uniform grid over $[1000, 1800]$\,Pa: 33 inlet pressures for the 7-edge network
and 20 for the tree network. Each run advances $T/\Delta t = 1000$
steps with $\Delta t = 0.01$\,s from the shut-in initial state,
storing the coefficient vector $c(t_n, P_{\mathrm{in}}) \in
\mathbb{R}^K$ at every step together with the basis matrix $\Phi$, the
weight vector $\mathrm{diag}(W)$, and the degree-of-freedom map that
records which rows and columns belong to each edge's pressure and
flux blocks. Held-out test pressures are excluded from this grid and their reference trajectories are generated by separate runs of the
same solver.

\paragraph{Latent basis.}
The stored coefficient trajectories are concatenated across all training inlet pressures into a snapshot matrix of size $(n_{P_{\mathrm{in}}} \cdot 1001) \times K$, mean-centred, and reduced by a thin singular value decomposition; the leading $r$ right singular vectors form $V_r$. The matrix includes the initial state at $t = 0$,
while the $1000$ subsequent states per trajectory form the training pairs. We use $r = 35$ for both networks, which retains more than
$99.99\%$ of the coefficient-space energy on each.

\paragraph{Initial-state coefficient.}
The shut-in state is uniform pressure $P_{\mathrm{in}}$ with zero
flux. We
compute it once as the least-squares projection
$c_1 = (\Phi^\top\Phi)^{-1}\Phi^\top e_P$ of the unit-pressure,
zero-flux grid state $e_P$ (ones on pressure degrees of freedom, zeros
on flux), and set $c_{\mathrm{init}}(P_{\mathrm{in}}) =
P_{\mathrm{in}}\, c_1$. The same expression is used in data generation
and inside the model, so the predicted state at $t = 0$ is consistent
with the reference by construction.

\paragraph{Precomputing the projected operators.}
For every training pair $(t_{n+1}, P_{\mathrm{in}})$ we reconstruct the
reference grid state $u^n = \Phi\, c_{\mathrm{ref}}(t_n,
P_{\mathrm{in}})$, assemble the discrete operator $A^n$ and right-hand
side $b^n$ from it exactly as in the solver, apply the row weights $W$,
and store
\begin{equation}
A_z^n = W A^n \Phi V_r \in \mathbb{R}^{m \times r},
\qquad
b_z^n = W\big(b^n - A^n \Phi\, c_{\mathrm{init}}(P_{\mathrm{in}})\big)
\in \mathbb{R}^{m},
\end{equation}
where $m$ is the number of rows of the discrete system (2811 for the
7-edge network, 4821 for the tree). This is done once, before
training, in single precision. The resulting tensors occupy
$13.0$\,GB and $13.5$\,GB respectively.

\paragraph{Training.}
$g_\theta$ is a three-layer perceptron (two hidden layers of width 128
with $\tanh$, linear output of width $r$) taking normalized time and
inlet pressure. Its output is scaled componentwise by
$z_{\mathrm{scale}}$, the standard deviation of the training latents
$V_r^\top(c_{\mathrm{ref}} - c_{\mathrm{init}})$ offset by one, which
puts all latent components on a comparable scale at initialization.
Training uses Adam at learning rate $10^{-3}$ with batches of 16
training pairs drawn in a fresh random permutation each epoch, and
gradients clipped at max-norm 20. Each batch evaluates the residual by
a batched matrix-vector product $A_z^n \delta z$, so no operator is
ever assembled during training. We keep the checkpoint with the lowest
epoch-mean training residual. Full settings are in
Table~\ref{tab:nbmol_hparams}.

\paragraph{Evaluation.}
At a queried $(t, P_{\mathrm{in}})$ the model returns
$c_{\mathrm{pred}} = c_{\mathrm{init}}(P_{\mathrm{in}}) + V_r\,\delta z$
in one forward pass; the grid state follows as $\Phi c_{\mathrm{pred}}$.
Relative $L^2$ errors are computed separately for pressure and flux by
selecting the corresponding degree-of-freedom indices from the
reconstructed state and comparing against the NBM reference trajectory
at the same inlet pressure over all 1000 stored times.

Table~\ref{tab:nbmol_hparams} lists the training configuration of
NBM-OL for both networks.

\begin{table}[h]
\centering
\caption{NBM-OL training hyperparameters.}
\label{tab:nbmol_hparams}
\begin{tabular}{lcc}
\toprule
 & 7-edge network & Tree network \\
\midrule
Training range of $P_{\mathrm{in}}$ & $[1000, 1800]$ & $[1000, 1800]$ \\
Training inlet pressures $n_{P_{\mathrm{in}}}$ & 33 (spacing 25) & 20 (spacing $\approx 42.1$) \\
Time steps per trajectory & 1000 & 1000 \\
Training pairs $(t_{n+1}, P_{\mathrm{in}})$ & 33{,}000 & 20{,}000 \\
NBM coefficients $K$ & 564 & 969 \\
Latent dimension $r$ & 35 & 35 \\
Stored latent operators $A_z^n$ & 13.0\,GB & 13.5\,GB \\
\midrule
MLP $g_\theta$ & \multicolumn{2}{c}{2 hidden layers, width 128, $\tanh$} \\
Latent scaling $z_{\mathrm{scale}}$ & \multicolumn{2}{c}{std.\ of training latents $+\,1$} \\
Optimizer & \multicolumn{2}{c}{Adam, initial learning rate $10^{-3}$} \\
Batch size & \multicolumn{2}{c}{16} \\
Epochs & 3000 & 2000 \\
LR schedule (ReduceLROnPlateau, factor 0.5) & patience 500, min $10^{-5}$ & patience 300, min $10^{-6}$ \\
Gradient clipping (max norm) & \multicolumn{2}{c}{20} \\
Checkpoint selection & \multicolumn{2}{c}{lowest training residual/latest} \\
\bottomrule
\end{tabular}
\end{table}

\section{Additional Results}
\label{app:results}

\subsection{Tree Network Verification}
\label{app:verify_tree}

\begin{figure}[!htbp]
\centering
\begin{subfigure}[b]{0.75\textwidth}
    \centering
    \includegraphics[width=\linewidth]{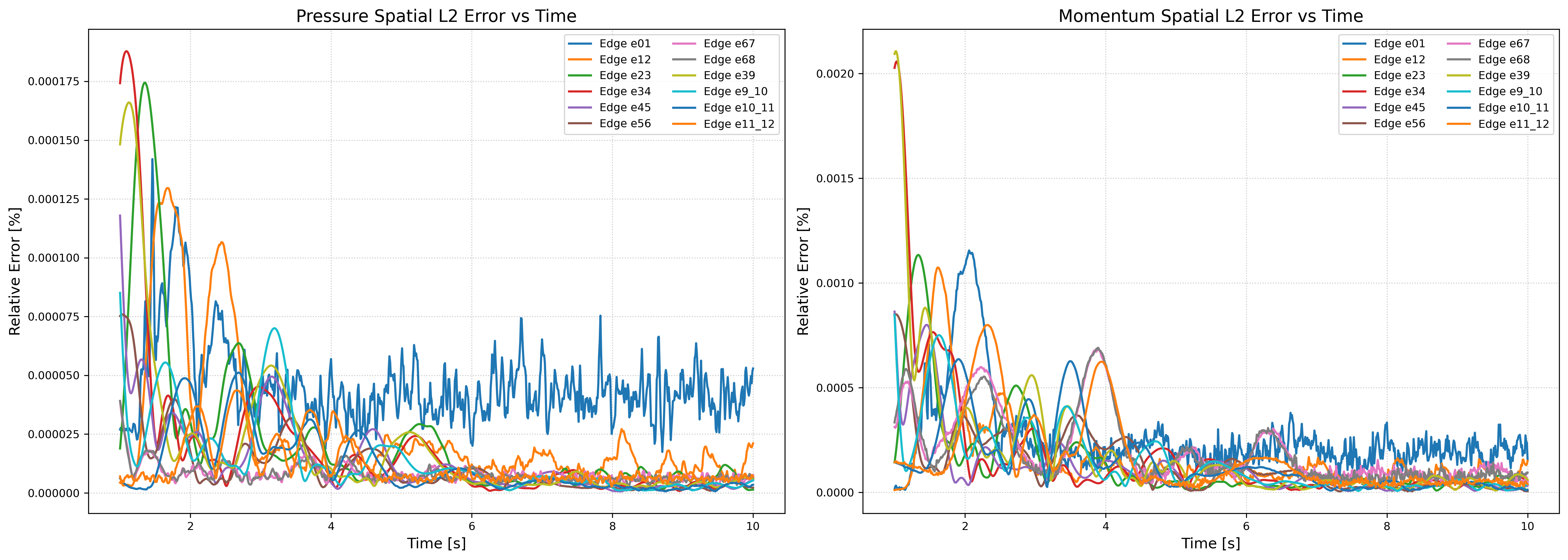}
    \caption{NBM vs.\ Finite Volume}
\end{subfigure}

\begin{subfigure}[b]{0.75\textwidth}
    \centering
    \includegraphics[width=\linewidth]{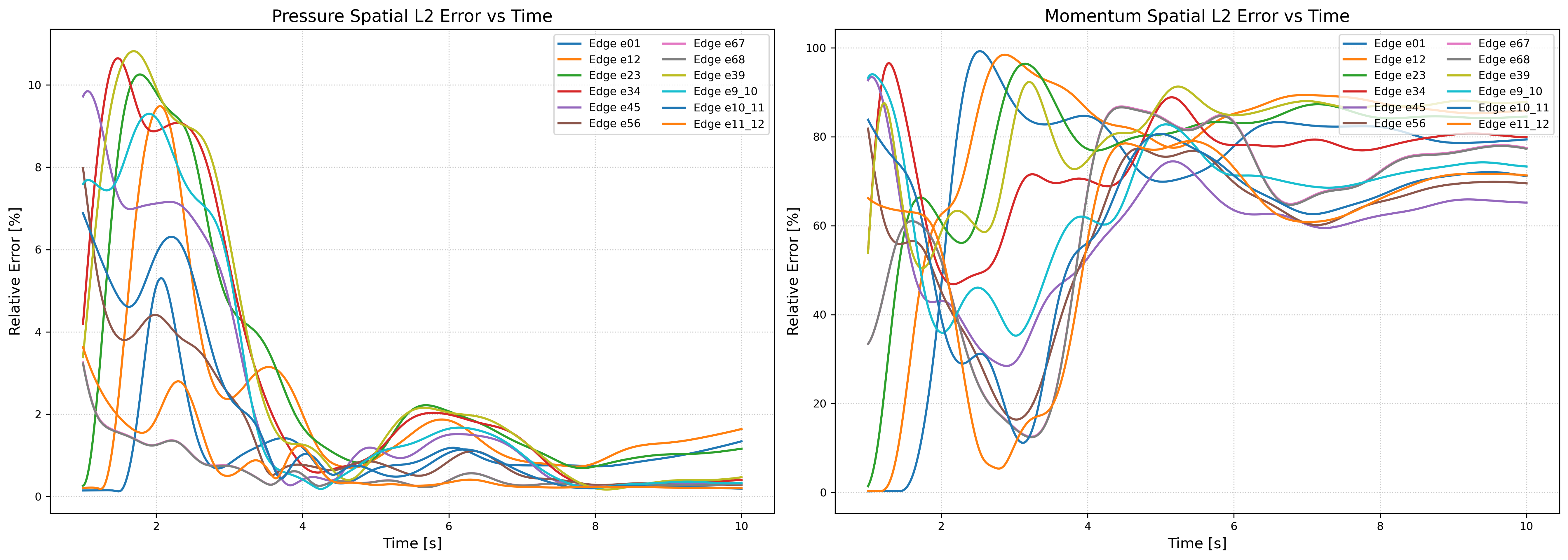}
    \caption{PINN vs.\ Finite Volume}
\end{subfigure}
\caption{Per-edge relative $L^2$ error over time, tree network,
against the finite-volume reference: (a) NBM, (b) PINN.}
\label{fig:belgian_l2-vs-time}
\end{figure}

Figure~\ref{fig:belgian_l2-vs-time} shows the per-edge relative $L^2$ error over time for the tree network, the counterpart of Figure~\ref{fig:7edge_l2-vs-time}; both are plotted from $t = 1$\,s,
after the initial expansion wave has crossed the network. NBM remains
below $2\times10^{-4}\%$ in pressure and $2\times10^{-3}\%$ in flux on
every edge, decaying to roughly $5\times10^{-5}\%$ and
$2\times10^{-4}\%$ once the transient passes. The PINN errors are four
to five orders of magnitude larger, and the two variables behave
differently: pressure error peaks near $11\%$ around $t = 2$\,s and
then decays to below $2\%$, whereas flux error stays between $60\%$
and $100\%$ on every edge for the entire horizon. The PINN therefore
recovers the pressure field approximately while never establishing the
throughflow, the same failure mode seen on the 7-edge network.

\subsection{Per-Point Operator-Learning Errors}
\label{app:ol_perpoint}

Tables~\ref{tab:holdout_interp}--\ref{tab:holdout_extrap_tree} report
the relative $L^2$ error at each held-out inlet pressure; their means appear in Table~\ref{tab:ol_main}. Each entry is computed over all edges, spatial points, and time steps of a single trajectory, obtained from one forward pass of NBM-OL per queried time. Three patterns are visible across the four tables. First, NBM-OL's
pressure error varies little within the training range, staying between $0.048\%$ and $0.081\%$ on the 7-edge network and between
$0.097\%$ and $0.145\%$ on the tree network, and grows outside it,
most steeply at $1880$\,Pa on the tree network ($1.69\%$). Second, its
flux error is largest at the lowest test pressures, $1.32\%$ at
$1215$\,Pa and $7.15\%$ at $1063.2$\,Pa, and falls as
$P_{\mathrm{in}}$ increases before rising again slightly at the top of
the range; the smallest driving pressure difference
$P_{\mathrm{in}} - P_{\mathrm{out}}$ occurs at the lowest test point,
so the same absolute error there corresponds to a larger relative one.
Third, PI-DeepONet shows the opposite trend in pressure: its error
grows monotonically with $P_{\mathrm{in}}$, from $5.24\%$ to $10.57\%$
on the 7-edge network and from $6.49\%$ to $17.74\%$ on the tree
network, while its flux error stays near saturation at $85$--$96\%$ at
every point in both regimes. Its pressure predictions are constrained
to $[P_{\mathrm{out}}, 1.2\,P_{\mathrm{in}}^{\max}]$ by construction
(Appendix~\ref{app:pideeponet}), so the growth with
$P_{\mathrm{in}}$ reflects an increasing deviation within that
interval rather than a boundary-condition violation. The tree network's test pressures each lie midway between two
training pressures on the $20$-point grid, so no interpolation point
coincides with a training value.

\begin{table}[h]
\centering
\caption{Interpolation error (\%) at each held-out inlet pressure,
7-edge network (training range $[1000, 1800]$\,Pa).}
\label{tab:holdout_interp}
\begin{tabular}{c cc cc}
\toprule
& \multicolumn{2}{c}{NBM-OL} & \multicolumn{2}{c}{PI-DeepONet} \\
\cmidrule(lr){2-3} \cmidrule(lr){4-5}
$P_{\text{in}}$ & Pressure & Flux & Pressure & Flux \\
\midrule
1215 & $\bp{0.0483}$ & $\bp{1.3165}$ & $5.2368\%$ & $89.4339\%$ \\
1365 & $\bp{0.0574}$ & $\bp{0.9894}$ & $6.6521\%$ & $88.0012\%$ \\
1415 & $\bp{0.0605}$ & $\bp{0.9398}$ & $7.0646\%$ & $87.5787\%$ \\
1715 & $\bp{0.0772}$ & $\bp{0.9546}$ & $9.9673\%$ & $85.7612\%$ \\
1765 & $\bp{0.0811}$ & $\bp{1.0283}$ & $10.5748\%$ & $85.5879\%$ \\
\midrule
Mean & $\bp{0.0649}$ & $\bp{1.0457}$ & $7.8991\%$ & $87.2726\%$ \\
\bottomrule
\end{tabular}
\end{table}

\begin{table}[h]
\centering
\caption{Extrapolation error (\%) at each held-out inlet pressure
above the training range, 7-edge network.}
\label{tab:holdout_extrap}
\begin{tabular}{c cc cc}
\toprule
& \multicolumn{2}{c}{NBM-OL} & \multicolumn{2}{c}{PI-DeepONet} \\
\cmidrule(lr){2-3} \cmidrule(lr){4-5}
$P_{\text{in}}$ & Pressure & Flux & Pressure & Flux \\
\midrule
1840 & $\bp{0.1253}$ & $\bp{1.2055}$ & $11.5568\%$ & $85.3996\%$ \\
1880 & $\bp{0.3354}$ & $\bp{2.0254}$ & $12.1181\%$ & $85.3338\%$ \\
\midrule
Mean & $\bp{0.2304}$ & $\bp{1.6154}$ & $11.8374\%$ & $85.3667\%$ \\
\bottomrule
\end{tabular}
\end{table}

\begin{table}[h]
\centering
\caption{Interpolation error (\%) at each held-out inlet pressure,
tree network (training range $[1000, 1800]$\,Pa). Each test pressure
lies midway between two training pressures.}
\label{tab:holdout_interp_tree}
\begin{tabular}{c cc cc}
\toprule
& \multicolumn{2}{c}{NBM-OL} & \multicolumn{2}{c}{PI-DeepONet} \\
\cmidrule(lr){2-3} \cmidrule(lr){4-5}
$P_{\text{in}}$ & Pressure & Flux & Pressure & Flux \\
\midrule
1063.2 & $\bp{0.1230}$ & $\bp{7.1528}$ & $6.4937\%$ & $96.4810\%$ \\
1231.6 & $\bp{0.0973}$ & $\bp{3.6356}$ & $10.4083\%$ & $96.0434\%$ \\
1400.0 & $\bp{0.1139}$ & $\bp{1.9213}$ & $13.9584\%$ & $95.2578\%$ \\
1568.4 & $\bp{0.1296}$ & $\bp{1.5229}$ & $16.3645\%$ & $94.6084\%$ \\
1736.8 & $\bp{0.1454}$ & $\bp{2.0224}$ & $17.7417\%$ & $94.2131\%$ \\
\midrule
Mean & $\bp{0.1218}$ & $\bp{3.2510}$ & $12.9933\%$ & $95.3207\%$ \\
\bottomrule
\end{tabular}
\end{table}

\begin{table}[h]
\centering
\caption{Extrapolation error (\%) at each held-out inlet pressure
above the training range, tree network.}
\label{tab:holdout_extrap_tree}
\begin{tabular}{c cc cc}
\toprule
& \multicolumn{2}{c}{NBM-OL} & \multicolumn{2}{c}{PI-DeepONet} \\
\cmidrule(lr){2-3} \cmidrule(lr){4-5}
$P_{\text{in}}$ & Pressure & Flux & Pressure & Flux \\
\midrule
1820 & $\bp{0.2374}$ & $\bp{2.3682}$ & $17.9332\%$ & $94.0869\%$ \\
1880 & $\bp{1.6927}$ & $\bp{3.3212}$ & $17.8690\%$ & $94.0200\%$ \\
\midrule
Mean & $\bp{0.9650}$ & $\bp{2.8447}$ & $17.9011\%$ & $94.0534\%$ \\
\bottomrule
\end{tabular}
\end{table}

\subsection{Tree Network Extrapolation}
\label{app:tree_extrap}

\begin{figure}[!htbp]
\centering
\begin{subfigure}[b]{\linewidth}
    \includegraphics[width=\linewidth]{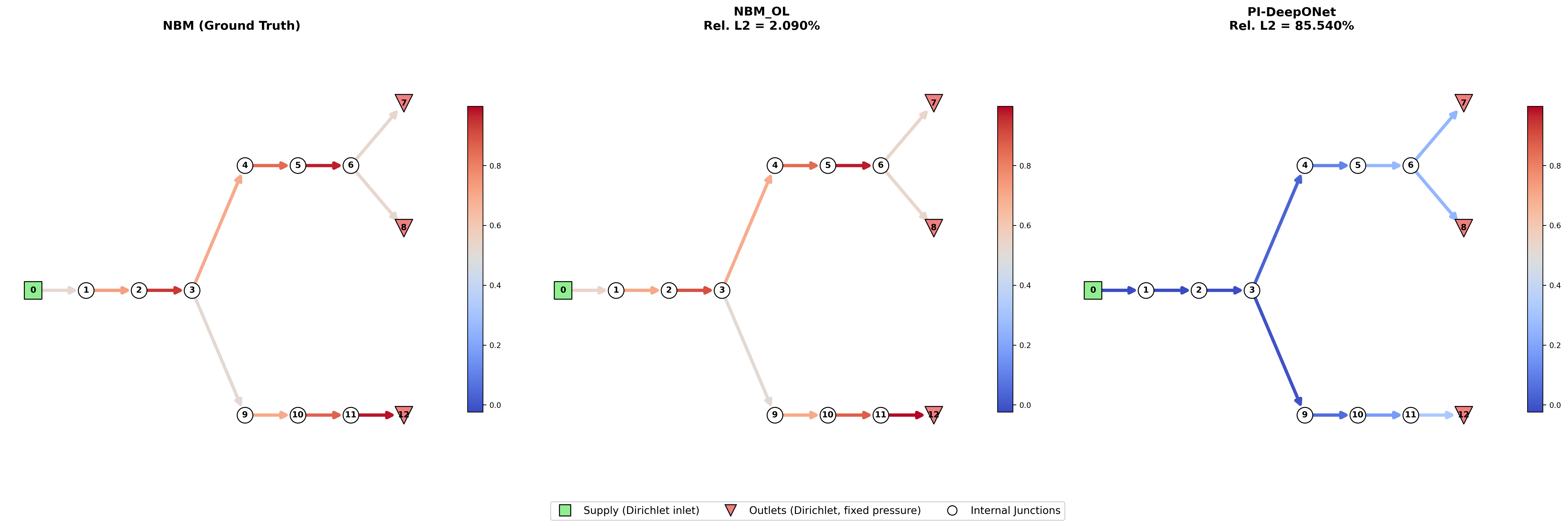}
    \caption{Network heatmap, momentum $q$ at $t=2.0$s}
\end{subfigure}
\vspace{4pt}
\begin{subfigure}[b]{\linewidth}
    \includegraphics[width=\linewidth]{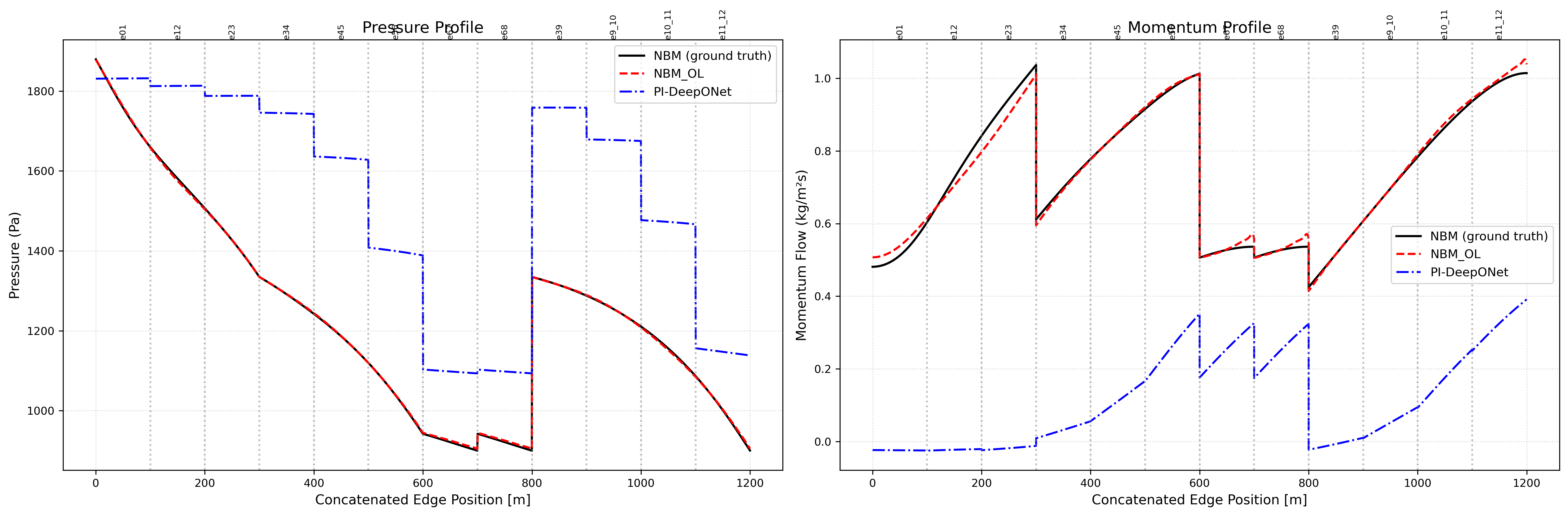}
    \caption{Spatial profile at $t=2.0$s}
\end{subfigure}
\vspace{4pt}
\begin{subfigure}[b]{\linewidth}
    \includegraphics[width=\linewidth]{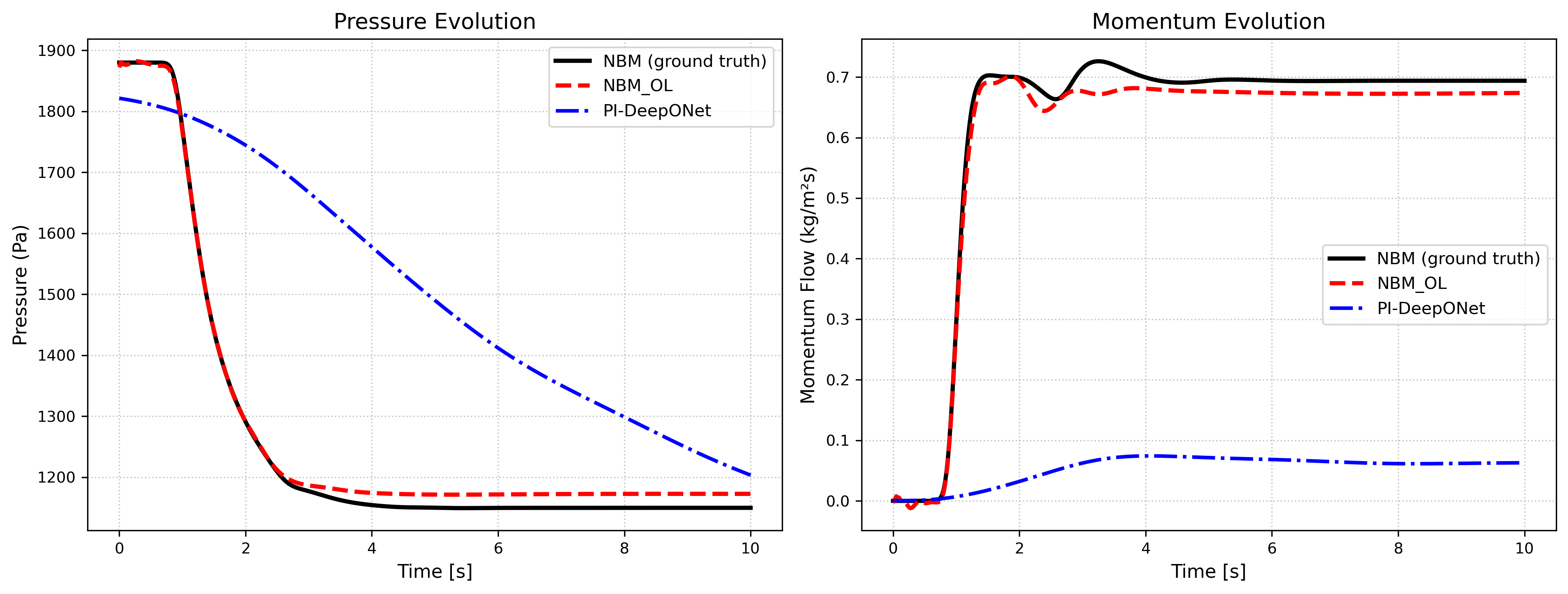}
    \caption{Time evolution, edge e34}
\end{subfigure}
\caption{Extrapolation, tree network, $P_{\mathrm{in}} = 1880$\,Pa
($10\%$ of the training range beyond its upper end). Reference NBM
(black), NBM-OL (red), PI-DeepONet (blue). (a) Flux over the network
at $t = 2$\,s. (b) Pressure and flux profiles along the concatenated
edges at $t = 2$\,s. (c) Time evolution at the midpoint of edge e34.}
\label{fig:extrap_tree_combined}
\end{figure}

Figure~\ref{fig:extrap_tree_combined} shows the tree network at
$P_{\mathrm{in}} = 1880$\,Pa, $10\%$ of the training range beyond its
upper end, the same case as Figure~\ref{fig:extrap_7edge_combined} for
the 7-edge network. NBM-OL reproduces the pressure drop along the
inlet chain and across both branches, and reaches the correct
steady-state flux on each outlet branch; the visible discrepancy is a
small steady-state offset on edge e34, where the predicted flux
settles slightly below the reference. PI-DeepONet under-predicts the
flux throughout by close to an order of magnitude and produces a
pressure profile with spurious jumps at the branch points, consistent
with its $94\%$ mean flux error in Table~\ref{tab:holdout_extrap_tree}.

\section{Physics-Informed Neural Network Baseline}
\label{app:pinn}

We compare against a physics-informed neural network (PINN), trained
at a fixed boundary condition ($P_{\text{in}}=1000$, $P_{\text{out}}=900$)
matching the finite-volume reference and NBM solution used in the
single-instance benchmark. The PINN follows the standard formulation
of \citet{raissi2019physics}: a fully-connected network trained by
minimizing a weighted sum of PDE-residual, boundary-condition,
initial-condition, and junction-coupling losses via automatic
differentiation, with no solution data. Loss weights are held
fixed throughout training, consistent with \citet{strelow2023physics},
who find that adaptive loss-balancing schemes offer no benefit over
well-chosen constant weights for gas transport problems specifically.

\subsection{Architecture}
\label{subsec:pinn_arch}

The PINN baseline learns a single, network-wide map
$(\text{edge}, x, t) \mapsto (P, q)$, implemented as a feed-forward
multilayer perceptron. The edge is encoded as a one-hot vector of length
$|\mathcal{E}|$, concatenated with normalized position
$x/L \in [0,1]$ and normalized time $t/T_{\text{final}} \in [0,1]$,
giving an input of dimension $|\mathcal{E}| + 2$. The network has 4
hidden layers of width 128 with $\tanh$ activations, followed by a
linear output layer producing two raw scalars.

The raw outputs are mapped to physical pressure and momentum flow via
\begin{align}
    P &= P_{\text{out}} + (P_{\text{in}} - P_{\text{out}})\,
         \sigma(\hat{o}_1), \label{eq:pinn_pressure} \\
    q &= Q_{\text{ref}}\, \hat{o}_2, \label{eq:pinn_flow}
\end{align}
where $\hat{o}_1, \hat{o}_2$ are the two raw network outputs and
$\sigma(\cdot)$ is the logistic sigmoid. \eqref{eq:pinn_pressure}
constrains predicted pressure to the physical interval
$[P_{\text{out}}, P_{\text{in}}]$ by construction. $Q_{\text{ref}}$ in
\eqref{eq:pinn_flow} is not a hand-tuned constant, but derived
directly from the problem's physical scales (Section~\ref{subsec:pinn_residual}).

\subsection{PDE Residual}
\label{subsec:pinn_residual}

For collocation points sampled per edge, spatial and temporal
derivatives of $P$ and $q$ are obtained via automatic differentiation
with respect to the normalized inputs, rescaled by $1/L$ and
$1/T_{\text{final}}$ respectively to recover physical-unit derivatives.
The residuals of \eqref{eq:mass_pq}--\eqref{eq:momentum_pq} are
evaluated directly at each collocation point, with pressure clamped to
a small positive floor before entering the advective and frictional
terms, consistent with the singularity discussed in
Section~\ref{subsec:pq_transform}.

The raw residuals are non-dimensionalized before entering the loss.
Writing $P = P_{\text{ref}}\hat{P}$, $q = Q_{\text{ref}}\hat{q}$,
$x = L\hat{x}$, $t = T_{\text{final}}\hat{t}$ and substituting into the
mass equation, the scale $Q_{\text{ref}}$ that makes the resulting
advective coefficient equal to $1$ is fixed by the other,
already-physical scales:
\begin{equation}
    Q_{\text{ref}} = \frac{P_{\text{ref}}\, L}{a^2\, T_{\text{final}}}.
    \label{eq:q_ref}
\end{equation}
Dividing the raw mass and momentum residuals by $P_{\text{ref}}/T_{\text{final}}$
and $Q_{\text{ref}}/T_{\text{final}}$ respectively is equivalent to
substituting the hat variables into the PDE directly, without
separately hardcoding the resulting dimensionless groups.

One resulting dimensionless group, $(a\,T_{\text{final}}/L)^2 \approx 1225$
for this problem's parameters, is intrinsically large: the acoustic wave
transit distance ($a\,T_{\text{final}} \approx 3500\,\text{m}$) far
exceeds the domain length ($100\,\text{m}$), so the pressure-gradient
term dominates the momentum residual even after correct
nondimensionalization. This reflects the problem's genuinely fast wave
dynamics rather than a scaling artifact, and the momentum residual is
correspondingly harder to drive down than the mass residual throughout
training.

\subsection{Loss Terms}
\label{subsec:pinn_loss}

In addition to the PDE residual, three supervised loss terms enforce
the network formulation of Section~\ref{subsec:network}. All
pressure-mismatch terms are normalized by $P_{\text{in}}^2$ and all
flow-mismatch terms by $Q_{\text{ref}}^2$, bringing them to the same
numeric scale as the dimensionless PDE residual; without this
normalization, a fixed loss weight cannot make progress against raw,
unnormalized-scale terms, and training collapses to the trivial
solution $q \equiv 0$, $P \equiv \text{const}$.

\textbf{Boundary loss.} For every edge endpoint incident to a Dirichlet
vertex $k$ with prescribed pressure $P_k$,
\begin{equation*}
    \mathcal{L}_{\text{bc}} = \sum_{k \in \mathcal{V}_{\text{Dirichlet}}}
        \; \sum_{e \,:\, k \in e} \;
        \mathbb{E}\left[\left(\frac{P_e(x_k, t) - P_k}{P_{\text{in}}}\right)^2\right].
\end{equation*}

\textbf{Initial-condition loss.} At $t=0$, the network is driven toward
the shut-in state (uniform pressure at $\max_k P_k$, zero flow):
\begin{equation*}
    \mathcal{L}_{\text{ic}} = \sum_{e \in \mathcal{E}} \mathbb{E}\left[
        \left(\frac{P_e(x, 0) - P_{\text{shut-in}}}{P_{\text{in}}}\right)^2
        + \left(\frac{q_e(x, 0)}{Q_{\text{ref}}}\right)^2 \right].
\end{equation*}

\textbf{Junction loss.} At every internal vertex $k$, pressure
continuity is enforced by penalizing deviation from the mean pressure
across incident edges, and mass conservation
(\eqref{eq:kirchhoff}) is enforced directly as a soft penalty:
\begin{equation*}
    \mathcal{L}_{\text{junction}} = \sum_{k} \left[
        \sum_{e \,:\, k \in e} \mathbb{E}\left[\left(\frac{P_e(x_k,t) - \bar{P}_k(t)}{P_{\text{in}}}\right)^2\right]
        \;+\; \mathbb{E}\left[\left(\frac{\sum_{e:e\to k} q_e(L_e,t) - \sum_{e:k\to e} q_e(0,t)}{Q_{\text{ref}}}\right)^2\right]
    \right].
\end{equation*}

The total training loss is a weighted sum,
\begin{equation*}
    \mathcal{L} = w_{\text{pde}}\,\mathcal{L}_{\text{pde}}
        + w_{\text{bc}}\,\mathcal{L}_{\text{bc}}
        + w_{\text{ic}}\,\mathcal{L}_{\text{ic}}
        + w_{\text{junction}}\,\mathcal{L}_{\text{junction}},
\end{equation*}
with $w_{\text{bc}} = w_{\text{ic}} = w_{\text{junction}} = 10$ and
$w_{\text{pde}} = 1$ throughout.

\subsection{Training Procedure}
\label{subsec:pinn_training}

The network is trained for 20,000 epochs
with the Adam optimizer, learning rate $10^{-3}$ decayed via
ReduceLROnPlateau (factor $0.5$, patience $1000$), gradient clipping at
max norm $20.0$, and loss weights $w_{\text{pde}}=1.0$,
$w_{\text{bc}}=w_{\text{ic}}=w_{\text{junction}}=10.0$. Collocation points are resampled
independently every epoch for the PDE residual, boundary, initial, and
junction terms.

\subsection{Network Instantiations}
\label{subsec:pinn_instantiations}

The PINN architecture, loss formulation, and training procedure are
applied unmodified to both network topologies; only the topology-derived
quantities (edge count, Dirichlet vertex set, and the resulting junction
structure) differ, as summarized in
Table~\ref{tab:fv_networks}.
All other hyperparameters (architecture size, loss weights, optimizer
settings) are held identical across both networks, summarized in
Table~\ref{tab:pinn_hyperparams}.

\begin{table}[h]
\centering
\caption{PINN training hyperparameters, identical across both network
topologies.}
\label{tab:pinn_hyperparams}
\begin{tabular}{ll}
\toprule
Hyperparameter & Value \\
\midrule
Hidden layers & 4 \\
Hidden width & 128 \\
Activation & $\tanh$ \\
Optimizer & Adam \\
Learning rate & $10^{-3}$ \\
LR schedule & ReduceLROnPlateau (factor 0.5, patience 1000, min $10^{-5}$) \\
Gradient clipping (max norm) & 20 \\
Collocation points per term & 256 \\
Loss weights ($w_{\text{pde}}, w_{\text{bc}}, w_{\text{ic}}, w_{\text{junction}}$) & $1, 10, 10, 10$ \\
Training epochs & 20{,}000 \\
Random seed & 0 \\
\bottomrule
\end{tabular}
\end{table}

\subsection{PI-DeepONet Baseline: Architecture and Training Details}
\label{app:pideeponet}

We benchmark against a physics-informed DeepONet (PI-DeepONet), adapted
from the collocation-based training paradigm of \citet{wang2021learning}
to the graph-structured pipeline networks. PI-DeepONet is trained with no supervised trajectory data: all
training signal comes from PDE, boundary-condition, initial-condition,
and junction residuals evaluated via automatic differentiation at
collocation points resampled every epoch. The same architecture and
training procedure are applied to both network topologies without
modification; only the topology-derived quantities (edge count,
boundary-incident edges, junction count) differ, as summarized below.

\paragraph{Architecture.} PI-DeepONet uses a branch-trunk architecture
with latent width $p=64$. The branch network maps the (normalized)
scalar boundary condition $P_{\text{in}}$ to a $2p$-dimensional
embedding via a 3-layer MLP with hidden width 128 and $\tanh$
activations. The trunk network takes a concatenated input of
[edge one-hot ($\in \mathbb{R}^{|\mathcal{E}|}$), normalized spatial
coordinate $x/L$, normalized time $t/T_{\text{final}}$] and similarly
outputs a $2p$-dimensional embedding via a 3-layer MLP with hidden width
128. Branch and trunk embeddings are split into two $p$-dimensional
halves, combined via inner product plus a learned bias, and mapped to
pressure $P$ and flow $q$ outputs. The edge one-hot encoding is a
necessary adaptation of the original single-domain PI-DeepONet
formulation to a multi-edge network domain, since the trunk net must
distinguish which edge a given $(x,t)$ query refers to.

Pressure output is constrained via $P = P_{\text{out}} +
(P_{\text{in}}^{\max} - P_{\text{out}}) \cdot 1.2\,\sigma(\hat{P})$,
where $\hat{P}$ is the raw network output, $\sigma$ is the sigmoid
function, and $P_{\text{in}}^{\max}$ is the fixed upper end of the
training range (not the sampled $P_{\text{in}}$ of the current query,
which would give every training example a differently sized output
window and a $P_{\text{in}}$-dependent error). Flow output is given by
$q = Q_{\text{ref}}\,\hat{q}$, where $\hat{q}$ is the raw network
output and $Q_{\text{ref}}$ is derived from the problem's physical
scales rather than a hand-tuned constant, as detailed below.

\paragraph{Training objective.} The total loss is a weighted sum of
four residual terms, evaluated at collocation points resampled every
epoch:
\begin{equation*}
\mathcal{L} = w_{\text{pde}}\mathcal{L}_{\text{pde}}
  + w_{\text{bc}}\mathcal{L}_{\text{bc}}
  + w_{\text{ic}}\mathcal{L}_{\text{ic}}
  + w_{\text{junction}}\mathcal{L}_{\text{junction}},
\end{equation*}
with $w_{\text{bc}} = w_{\text{ic}} = w_{\text{junction}} = 10$ and
$w_{\text{pde}} = 1$.

\textbf{PDE residual.} Mass and momentum conservation in $(P,q)$ form,
evaluated at interior collocation points $(x,t) \sim U[0,1]^2$,
$P_{\text{in}} \sim U[P_{\text{in,min}}, P_{\text{in,max}}]$, for each
edge independently:
\begin{align*}
r_{\text{mass}} &= \partial_t P + a^2 \partial_x q, \\
r_{\text{mom}} &= \partial_t q + \partial_x P
  + \partial_x\!\left(\frac{a^2 q^2}{P}\right)
  + \frac{f a^2 |q| q}{2 D P}.
\end{align*}
The raw residuals are non-dimensionalized before entering the loss,
following the same derivation as the plain-PINN baseline: writing
$P = P_{\text{ref}}\hat{P}$, $q = Q_{\text{ref}}\hat{q}$, $x = L\hat{x}$,
$t = T_{\text{final}}\hat{t}$ and substituting into the mass equation,
the scale that makes the resulting advective coefficient equal to $1$
is
\begin{equation*}
    Q_{\text{ref}} = \frac{P_{\text{ref}}\, L}{a^2\, T_{\text{final}}},
\end{equation*}
with $P_{\text{ref}} = P_{\text{in}}^{\max}$, the same fixed constant
used for the pressure output constraint above. Dividing the raw
residuals by $P_{\text{ref}}/T_{\text{final}}$ and
$Q_{\text{ref}}/T_{\text{final}}$ respectively gives the
non-dimensionalized residuals $\hat{r}_{\text{mass}}, \hat{r}_{\text{mom}}$
entering the loss:
$\mathcal{L}_{\text{pde}} = \sum_{\text{edges}} \mathbb{E}[\hat{r}_{\text{mass}}^2 + \hat{r}_{\text{mom}}^2]$.
Without this normalization, the residual is dominated by the
coefficient $a^2 = 122500$, several orders of magnitude larger than the
already-normalized boundary, initial-condition, and junction terms
below, and training collapses to the trivial solution
$q \equiv 0$, $P \equiv \text{const}$, which trivially satisfies the
(unnormalized) interior residual while violating the boundary data. As
in the plain-PINN baseline, one resulting dimensionless group,
$(a\,T_{\text{final}}/L)^2 \approx 1225$, is intrinsically large for
this problem's parameters, so the momentum residual remains harder to
drive down than the mass residual even after correct
non-dimensionalization.

\textbf{Boundary condition loss.} At every edge endpoint incident to the
inlet vertex ($x=0$): $P = P_{\text{in}}$ (the sampled value). At every
edge endpoint incident to an outlet vertex ($x=1$): $P = P_{\text{out}}$
(fixed). A vertex with multiple incident edges contributes one
evaluation per edge, since each edge's prediction is an independent
network output with no shared degree of freedom across edges. Both
pressure-mismatch terms are normalized by $(P_{\text{in}}^{\max})^2$.

\textbf{Initial condition loss.} At $t=0$, all edges: $P = P_{\text{in}}$,
$q = 0$ (shut-in state, matching the NBM/FV initial condition exactly).
The pressure term is normalized by $(P_{\text{in}}^{\max})^2$ and the
flow term by $Q_{\text{ref}}^2$.

\textbf{Junction loss.} At each internal junction node: pressure
continuity across all incident edges, plus mass conservation
(net inflow $=$ net outflow). The pressure-continuity term is
normalized by $(P_{\text{in}}^{\max})^2$ and the mass-conservation term
by $Q_{\text{ref}}^2$.

\paragraph{Collocation points.} Table~\ref{tab:pideeponet_colloc}
summarizes the collocation points sampled per loss term, per epoch, for
each topology. All points are freshly resampled every epoch; no fixed
training set exists, so no discrete $P_{\text{in}}$ value is ever
repeated across training. Held-out interpolation points therefore lie
within the domain $[P_{\text{in,min}}, P_{\text{in,max}}]$ that the
physics losses were enforced over continuously, but were never seen as discrete supervised targets.

\begin{table}[h]
\centering
\caption{Collocation points sampled per training epoch for PI-DeepONet.
7-edge: 2 inlet-incident edges, 1 outlet-incident edge, 4 junctions.
Tree: 1 inlet-incident edge, 3 outlet-incident edges, 9 junctions.}
\label{tab:pideeponet_colloc}
\begin{tabular}{lcc}
\toprule
Loss term & 7-edge network & Tree network \\
\midrule
PDE residual & $256 \times 7 = 1792$ & $256 \times 12 = 3072$ \\
BC loss & $256 \times 3 = 768$ & $256 \times 4 = 1024$ \\
IC loss & $256 \times 7 = 1792$ & $256 \times 12 = 3072$ \\
Junction loss & $256 \times 4 = 1024$ & $256 \times 9 = 2304$ \\
\bottomrule
\end{tabular}
\end{table}

\paragraph{Hyperparameters.} Table~\ref{tab:pideeponet_hparams}
summarizes training hyperparameters, identical across both topologies.

\begin{table}[h]
\centering
\caption{PI-DeepONet training hyperparameters, identical across both
network topologies.}
\label{tab:pideeponet_hparams}
\begin{tabular}{ll}
\toprule
Hyperparameter & Value \\
\midrule
$P_{\text{in}}$ range & $[1000, 1800]$ \\
$P_{\text{out}}$ (fixed) & $900.0$ \\
$T_{\text{final}}$ & $10.0\,\text{s}$ \\
Epochs & $20{,}000$ \\
Collocation points per instance ($n_{\text{colloc}}$) & $256$ \\
Latent width $p$ & $64$ \\
Branch / trunk hidden width & $128$ \\
Network depth & $3$ layers \\
Optimizer & Adam \\
Learning rate & $1\times10^{-3}$ \\
LR schedule & ReduceLROnPlateau (factor 0.5, patience 1000, min $10^{-6}$) \\
Gradient clipping (max norm) & 20 \\
Loss weights ($w_{\text{pde}}, w_{\text{bc}}, w_{\text{ic}}, w_{\text{junction}}$) & $1, 10, 10, 10$ \\
Seed & $0$ \\
\bottomrule
\end{tabular}
\end{table}
\end{document}